\documentclass[journal]{IEEEtran}

\newcommand{\mc}[1]{\mathcal{#1}}
\newcommand{\bs}[1]{\boldsymbol{#1}}
\newcommand{\sls}[1]{\backslash \{#1\}}
\newcommand{\fa}[2]{\, \forall \, #1 \in \mathcal{#2}}

\usepackage{amssymb,amsmath,color,graphicx}
\usepackage{subfigure}
\usepackage{cite}
\usepackage{verbatim}
\usepackage{algorithm}
\usepackage{algcompatible} 
\usepackage{url}
\usepackage{graphicx} 

\DeclareMathOperator*{\argmax}{arg\,max}

\newtheorem{theorem}{Theorem}
\newtheorem{assumption}{Assumption}

\newtheorem{definition}{Definition}

\newtheorem{lemma}{Lemma}

\begin{document}

\title{Congestion-Aware Distributed Network Selection \\ for Integrated Cellular and Wi-Fi Networks}

\author{Man Hon Cheung, Fen Hou, \IEEEmembership{Member,~IEEE}, Jianwei Huang,~\IEEEmembership{Fellow,~IEEE}, and Richard Southwell
\thanks{
This work is supported by the Macau Science and Technology Development Fund under grant FDCT 121/2014/A3 and the Research Committee of University of Macau under grant MYRG2016-00171-FST. This work is also supported by the General Research Funds (Project Number CUHK 14202814, 14206315, and 14219016) established under the University Grant Committee of the Hong Kong Special Administrative Region, China.
 Part of this paper was presented in \emph{CISS'14} \cite{cheung_ca14}.}
\thanks{M.~H.~Cheung is affiliated with both the University of Macau and the Chinese University of Hong Kong; E-mail: mhcheung@ie.cuhk.edu.hk.
F.~Hou is with the Department of Electrical and Computer Engineering, University of Macau, Macau; E-mail: fenhou@umac.mo.
J.~Huang is with the Network Communications and Economics Lab (NCEL), Department of Information Engineering, the Chinese University of Hong Kong, Hong Kong, China; E-mail: jwhuang@ie.cuhk.edu.hk.
R.~Southwell is with the Department of Mathematics, University of York, York, United Kingdom; E-mail: richard.southwell@york.ac.uk.}
}


\maketitle

\thispagestyle{empty}

\begin{abstract}
  Intelligent network selection plays an important role in achieving an effective data offloading in the integrated cellular and Wi-Fi networks.
  However, previously proposed network selection schemes mainly focused on offloading as much data traffic to Wi-Fi as possible, without systematically considering the \emph{Wi-Fi network congestion} and the \emph{ping-pong effect}, both of which may lead to a poor overall user quality of experience.
  Thus, in this paper, we study a more practical network selection problem by considering both the impacts of the \emph{network congestion} and \emph{switching penalties}. 
  More specifically, we formulate the users' interactions as a Bayesian network selection game (NSG) under the incomplete information of the users' mobilities. We prove that it is a Bayesian potential game and show the existence of a pure Bayesian Nash equilibrium that can be easily reached. We then propose a \emph{distributed} network selection (DNS) algorithm based on the network congestion statistics obtained from the operator.
	Furthermore, we show that computing the optimal \emph{centralized} network allocation is an NP-hard problem, which further justifies our distributed approach.
Simulation results show that the DNS algorithm achieves the highest user utility and a good fairness among users, as compared with the on-the-spot offloading and cellular-only benchmark schemes.
\end{abstract}


\normalsize

\section{Introduction} \label{sec:intro}

 With the proliferation of mobile video and web applications, consumer demands for wireless data services are growing rapidly, to a point that the cellular network capacity is pushed to its limit.
  According to Cisco's forecast, mobile data traffic will increase to $30.6$ exabytes per month by 2020, which corresponds to an $8$-fold increase between 2015 and 2020 globally \cite{cisco_cv16}. The cellular network capacity, however, is growing at a much slower pace, and cannot keep up with the explosive growth in data traffic.
	A cost-effective and timely solution for alleviating the cellular network congestion problem is to use some complementary technologies, such as Wi-Fi or small cells, to offload the cellular traffic. 
	In fact, Cisco predicted that the percentage of offloaded traffic will grow and exceed that of the cellular traffic, reaching an offloading ratio of 55\% of the global mobile traffic  by 2020 \cite{cisco_cv16}.	
  Owing to the existing popularity of Wi-Fi usage and deployment\footnote{From Cisco's data, $64.2$ million public Wi-Fi hotspots have already installed since 2015 \cite{cisco_cv16}.}, we will focus on the network selection in the integrated cellular and Wi-Fi networks in this paper.

  Through the current ongoing standardization efforts, such as the access network discovery and selection function (ANDSF) and Hotspot 2.0 \cite{lucent_wr12, 4gamericas_io13}, the cellular and Wi-Fi networks are becoming more tightly coupled with each other. More specifically, under this cellular and Wi-Fi integration, the Wi-Fi networks would usually be owned and managed by the cellular operator, who ensures a seamless connectivity for the users. Also, the same operator will be responsible for making all the network selections silently in the background, so a user does not need to know whether he is connected to the cellular or Wi-Fi network. Furthermore, the operator will ensure that all functionality and services are consistently available regardless of whether the user is on cellular or Wi-Fi. 

  \emph{Intelligent network selection} plays a critical role in the integrated cellular and Wi-Fi networks, to achieve an effective mobile data offloading and improve the users' quality of experience (QoE).  One popular choice that is used by many smartphones by default is the on-the-spot offloading (OTSO) scheme, where the device simply offloads its data traffic to a Wi-Fi network whenever possible, and only uses the cellular network if no Wi-Fi exists (or the Wi-Fi interface is turned off).	
	The OTSO scheme is simple to implement but has two possible drawbacks.
  First, under the OTSO policy, devices that are in close proximity may choose the same Wi-Fi network, hence experience the \emph{network congestion} and achieve low throughput, especially during the peak hours in some densely populated areas. 
	In other words, this users' \emph{herd behaviour} without any \emph{coordination} leads to the Wi-Fi network congestion \cite{paolini_tw12}. 
  Second, a user may incur a \emph{switching penalty} in the forms of switching time and a switching cost when it switches between different networks. The switching time corresponds to the delay during handoff, and the switching cost accounts for the additional power consumption and QoE disruption \cite{southwell_sm12}. 
  Without taking into account this switching penalty, a network selection policy may result in the \emph{ping-pong effect} \cite{4gamericas_io13} with too frequent network switching, which leads to a throughput reduction and faster battery degradation.

  Although \emph{network congestion} and \emph{switching penalty} are two important factors in the design of an effective \emph{intelligent network selection} algorithm, most prior related literature, including \cite{balasubramanian_am10, lee_md10, ristanovic_ee11, im_ae13, moon_pd15}, neglected the effects of these two factors. 
  Balasubramanian \emph{et al.} in \cite{balasubramanian_am10} proposed that a user can perform data offloading by making predictions of future Wi-Fi availability using the past mobility history.
  Lee \emph{et al.} in \cite{lee_md10} described the on-the-spot offloading (OTSO) scheme that most smartphones are using today by default. 
  Ristanovic \emph{et al.} in \cite{ristanovic_ee11} considered an energy-efficient offloading for delay-tolerant applications. They proposed to extract typical users' mobility profiles for the prediction of Wi-Fi availabilities. 
  Im \emph{et al.} in \cite{im_ae13} considered the cost-throughput-delay tradeoff in user-initiated Wi-Fi offloading. Given the predicted future usage and the availability of Wi-Fi, the proposed system decides which application should offload its traffic to Wi-Fi at a given time, while taking into account the cellular budget constraint of the user.  
  Moon \emph{et al.} in \cite{moon_pd15} implemented a new transport layer to handle network disruption and delay for the development of delay-tolerant Wi-Fi offloading apps, by scheduling multiple flows to meet their deadlines with the maximal Wi-Fi usage.	
	Furthermore, although the studies in \cite{aryafar_rs13, monsef_cp15, mahindra_ap14, hu_an16} took the network congestion into account, they did not consider the effect of the switching penalty. 
  Aryafar \emph{et al.} in \cite{aryafar_rs13} studied the network selection dynamics in heterogeneous wireless networks under two classes of throughput models. They characterized the Pareto-efficiency of the equilibria and proposed a network selection algorithm with hysteresis mechanism.
  Following the work in \cite{aryafar_rs13}, Monsef \emph{et al.} in \cite{monsef_cp15} first considered a client-centric network selection model for autonomous user decision, and characterized the convergence time and conditions. They further studied a hybrid client-network model, where a user is allowed to switch network if this decision is in line with the network controller's potential function. 
  Mahindra \emph{et al.} in \cite{mahindra_ap14} considered the practical implementation of the intelligent network selection in LTE and Wi-Fi networks. The system consists of an interface assignment algorithm that dynamically assigns user flows to interfaces and an interface switching service that performs seamless interface switching for HTTP-based flows.
	Hu \emph{et al.} in \cite{hu_an16} proposed an adaptive network selection algorithm based on the attractor selection mechanism for the users to dynamically select the suitable access points. Both the offloading effectiveness and traffic delay were considered as the performance metrics.
	In summary, the network selection problem considering the network congestion and switching penalty in data offloading has not been explored in the literature.
	
	In this paper, we jointly consider both the network congestion and switching penalty and address the practical considerations of user mobility, and location, user, and time dependent Wi-Fi availabilities.	
  For the user mobility, we assume that the operator only has the statistical information about the users' mobility patterns, which capture their daily movement habits \cite{ghosh_op06}. 
  We also consider several general assumptions on the Wi-Fi availabilities. First, we assume that the Wi-Fi availability is \emph{location-dependent}, because Wi-Fi access points (APs) are only available at some limited locations due to their smaller coverages. Second, it may be \emph{time-dependent} due to the access policies of the administrators of the Wi-Fi APs. For example, some Wi-Fi APs may be configured in the open access mode when the owner is away, but in the closed access mode when the owner is back. Third, it may be \emph{user-dependent}, as users who have subscribed to different data plans or Wi-Fi services (e.g., Skype Wi-Fi) can have different privileges to access different Wi-Fi networks. 
  Given these practical considerations with heterogeneous users and networks, the network selection problem is very challenging to tackle.

  Due to the coupling of the users' decisions in causing the network congestion, we apply the non-cooperative game theory to study this congestion-aware network selection problem. More specifically, with the statistical information on users' mobility patterns, we formulate the users' network selections over a period of multiple time slots as a \emph{Bayesian game} \cite{shoham_ma08}. In general, it is difficult to characterize the existence and convergence of the Bayesian Nash equilibrium.
	Nevertheless, we are able to show that the formulated game is a \emph{Bayesian potential game} \cite{facchini_cm97}, which enables us to design a distributed network selection (DNS) with some nice convergence properties. 
  It should be noted that convergence is important for congestion-aware network selections, where users switch networks based on the experienced network congestion levels. Without convergence guarantees, the system may result in oscillations.
	In addition, as a benchmark, we show that computing the socially optimal solution that maximizes the users' aggregate utilities in a centralized setting is an NP-hard problem.


  In summary, the main contributions of our work are as follows:
\begin{itemize}

\item \emph{Practical modeling}: We study the users' network selection problem by taking into account the practical issues of network congestion, switching penalties, and statistical information of the users' various possible mobility patterns.

\item \emph{NP-hard centralized network allocation benchmark}: We show that maximizing the users' aggregate utilities is an NP-hard problem, which motivates us to consider the distributed setting. 

\item \emph{Distributed network selection algorithm}: We formulate the users' network selection interactions as a Bayesian game. We show that it is a potential game, derive its closed-form exact potential function, and propose a practical DNS algorithm with nice convergence properties.

\item \emph{Load balancing}: Simulation results show that the proposed DNS scheme achieves a good fairness and improves the user utility of the cellular-only and OTSO schemes by $66.7\%$ under a medium switching cost.
  We also show that the OTSO scheme performs reasonably well with a low switching cost and a low Wi-Fi availability.

\end{itemize}

  The rest of the paper is organized as follows. 
	We first describe the system model in Section \ref{sec:model}. 
  We study the centralized network allocation and the distributed network selection game in Sections \ref{sec:centralized} and \ref{sec:game}, respectively.
  We present the simulation results in Section \ref{sec:pe} and conclude the paper in Section \ref{sec:concl}.

\section{System Model} \label{sec:model}
  
	In this section, we discuss the system model for the network selection in the integrated cellular and Wi-Fi networks.
  More specifically, we describe the networks setting in Section \ref{sec:network} and a user's  network availability and mobility pattern in Section \ref{sec:user}.
	We present his action as a network-time routes in Section \ref{sec:networktime} and his utility function in Section \ref{sec:utility}.

\begin{figure}[t]
 \centering
   \includegraphics[width=8cm, trim = 1cm 3cm 2cm 2cm, clip = true]{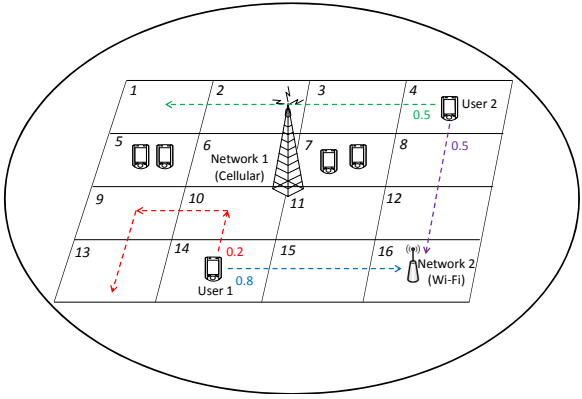} 
 \caption{An example of the integrated cellular and Wi-Fi network, where $\mathcal{N} = \{0, 1,2\}$ is the set of networks and $\mathcal{L} = \{1,\ldots,16\}$ is the set of locations. In each time slot, a user can remain idle (i.e., choose the auxiliary network $0$), access the cellular network (i.e., network $1$) or the Wi-Fi network (i.e., network $2$) if available, as the Wi-Fi availability is user, location, and time dependent. Each user may have \emph{multiple} possible mobility patterns, where the operator has \emph{incomplete} information on their probability distributions. We consider the users' network selections across a time period of multiple time slots by taking into account the \emph{user mobility}, \emph{network congestion}, and \emph{switching penalties}.} 
\label{fig:network}
\end{figure}

\subsection{Network Setting} \label{sec:network}

  As shown in Fig.~\ref{fig:network}, we consider an integrated cellular and Wi-Fi system, where the Wi-Fi networks are tightly integrated with the cellular network in terms of the radio frequency coordination and network management \cite{ericsson_wi12}.	
  Let $\mathcal{N} = \{0,1,\ldots,N\}$ be the set of $N + 1$ networks, where network $n = 1$ corresponds to the cellular network and network $n \in \mathcal{N}_{\text{wifi}} = \{2,\ldots,N\}$ corresponds to a Wi-Fi network.  We introduce an auxiliary \emph{idle network} $n = 0$ to model the situation that the user chooses to remain idle and is not actively using any networks.
  The network parameters are described as follows.

\begin{definition}[Network Parameters]
  Each network $n\in\mathcal{N}$ is associated with:	
\begin{itemize}
\item \emph{Network capacity} $\mu[n]$: The maximum total amount of data rate that network $n$ can serve the users in each time slot. 

\item \emph{Switching cost} $c[n,n']$: The cost incurred by a user when he switches from network $n \in \mathcal{N}$ to network $n' \in \mathcal{N}$. It can account for additional power consumption and QoE disruption \cite{southwell_sm12} during network switching. 

\item \emph{Switching time} $\delta[n,n']$: The delay incurred by a user when he switches from network $n \in \mathcal{N}$ to network $n' \in \mathcal{N}$, which is the total number of time slots required to tear down the old connection of network $n$ and setup the new connection of network $n'$. It corresponds to the delay during handoff between different wireless networks. 

\end{itemize}
\end{definition}

  To account for the fact that there is no network switching when a user keeps using network $n \in \mathcal{N}$, we have $c[n,n] = 0$ and $\delta[n,n] = 0, \, \forall \, n \in \mathcal{N}$.
  For the idle network, we make the following additional practical assumptions:

\begin{assumption}[Idle network] \label{ass:idle}
  (a) $\mu[0] = 0$;
	(b) The switching time through the idle network satisfies
\begin{equation} \label{equ:triangleineq_delta}
	\delta[n,n'] + \delta[n',n''] \geq \delta[n,0] + \delta[0,n''], \, \forall \, n,n',n'' \in \mathcal{N} \sls{0}.
\end{equation}
	(c) The switching cost through the idle network satisfies
\begin{equation} \label{equ:triangleineq}
	c[n,n'] + c[n',n''] > c[n,0] + c[0,n''], \, \forall \, n,n',n'' \in \mathcal{N} \sls{0}.
\end{equation}
\end{assumption}

 Assumption \ref{ass:idle}(a) implies that a user cannot receive any data during the idle state. 
 Assumption \ref{ass:idle}(b) accounts for the fact that an idle state requires less time to ``setup'' or ``tear down'' than switching through a third network $n'$.
 Assumption \ref{ass:idle}(c) captures the additional power and signalling overhead during handovers that involves one more network $n'$.\footnote{It is possible to use strict inequalities in both \eqref{equ:triangleineq_delta} and \eqref{equ:triangleineq}. However, since the switching time is an integer in this paper, it is more practical to consider an inequality in \eqref{equ:triangleineq_delta}.}

\subsection{User Setting} \label{sec:user}

  Let $\mathcal{I} = \{1,\ldots,I\}$ be the set of users, $\mathcal{L} = \{1,\ldots,L\}$ be the set of locations, and $\mathcal{T} = \{1,\ldots,T\}$ be the set of time slots. 
  We define a user's network availability\footnote{Our modeling on network availability is quite general as it allows each location to have more than one Wi-Fi access points (APs) and each AP to cover more than one location. Thus, there can be overlapping coverage areas of different networks. Also, it is a straightforward extension to consider multiple cellular networks (e.g., deployed by different mobile operators), and assum e that each location can be covered by an arbitrary number of these networks.} and mobility pattern as follows.

\begin{definition}[User's Network Availability and Mobility Pattern]  \label{def:attributes_user}
A user $i \in \mathcal{I}$ is associated with:
\begin{itemize}

\item \emph{User, location, and time dependent network availabilities} $\mathcal{M}[i,l,t] \subseteq \mathcal{N}$: The set of networks available for user $i \in \mathcal{I}$ at location $l \in \mathcal{L}$ and time $t \in \mathcal{T}$.\footnote{For the rest of the paper, we will assume that the idle network is available for all the users at all possible locations and time slots, so that network $0 \in \mathcal{M}[i,l,t], \, \forall \, i \in \mathcal{I}, l \in \mathcal{L}, t \in \mathcal{T}$.}

\item \emph{Mobility pattern} $\boldsymbol{\theta}_i = (l[i,t] \in \mathcal{L}, \, \forall \, t \in \mathcal{T}) \in \boldsymbol{\Theta}_i$: The locations of user $i$ in the period of $T$ time slots due to his mobility, where $l[i,t]$ is the position of user $i$ at time $t$, and $\boldsymbol{\Theta}_i$ is the set of all possible mobility patterns of user $i$ given his initial location at time $t = 1$.
  Each user may have \emph{multiple} mobility patterns. As an example, in Fig.~\ref{fig:network}, for a total of $T = 4$ slots, user $i$ has two possible mobility patterns: $\boldsymbol{\theta}_1^{(1)} = (14, 15, 16, 16)$ and $\boldsymbol{\theta}_1^{(2)} = (14, 10, 9, 13)$, so $\boldsymbol{\Theta}_1 = \{ \boldsymbol{\theta}_1^{(1)}, \boldsymbol{\theta}_1^{(2)} \}$. 
  We also refer to $\boldsymbol{\theta}_i$ as the \emph{type} of user $i$.
	
\item \emph{Prior Probability on Mobility Pattern} $p(\boldsymbol{\theta}_i) \geq 0$: The probability that user $i$ chooses mobility pattern $\boldsymbol{\theta}_i$.\footnote{Thus, $p(\boldsymbol{\theta}_i)$ is a system parameter on user $i$'s mobility, which can be collected from the mobile device automatically in the background.} We have $\sum_{\bs{\theta_i} \in \Theta_i} p(\boldsymbol{\theta}_i) = 1$. 
  As an example in Fig.~\ref{fig:network}, we have $p(\boldsymbol{\theta}_1^{(1)}) = 0.8$ and $p(\boldsymbol{\theta}_1^{(2)}) = 0.2$.

\end{itemize}
\end{definition}	
	
  We refer to this general mobility setting as the \emph{random} mobility pattern case. It includes the special case of \emph{deterministic} mobility pattern, where the user knows his own mobility pattern accurately.
	
\begin{figure}[t]
 \centering
	 \includegraphics[width=6cm, trim = 6cm 3.5cm 6cm 2cm, clip = true]{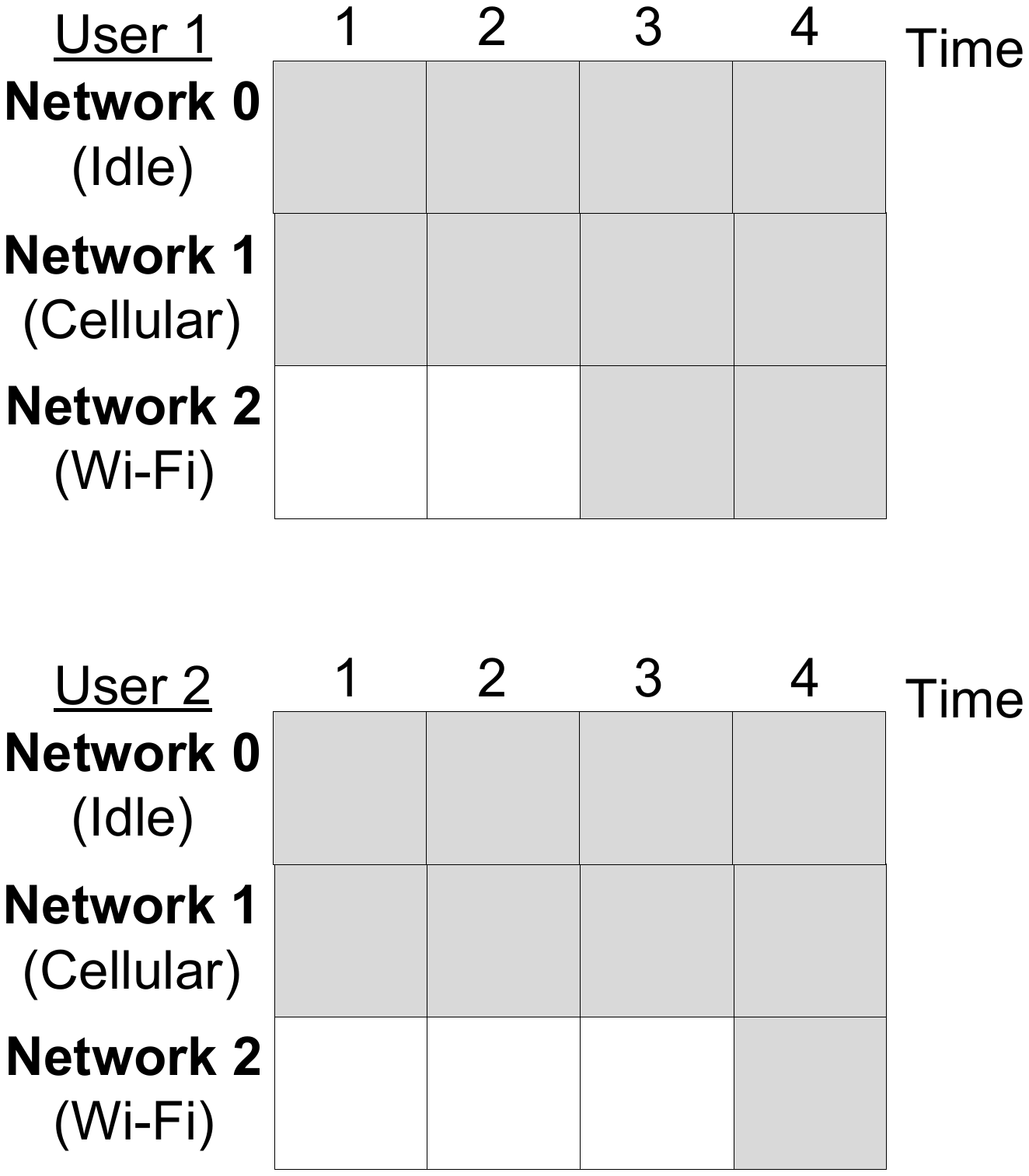} 
 \caption{Heterogeneous network availabilities of user 1 and user 2 shown in Fig.~\ref{fig:network} under mobility patterns $\bs{\theta}_1 = (14,15,16,16)$ and $\bs{\theta}_2 = (4,8,12,16)$. Here, a grey block means that the network is available (i.e., network $n \in \mathcal{N}[i,t]$) and a white block means that the network is not available.}
\label{fig:resourceblock}
\end{figure}

  Given user $i$'s mobility pattern and the network availabilities, we can compute his available set of networks at time $t$ as\footnote{As a result, we will focus on $\mathcal{N}[i,t]$, instead of $\mathcal{M}[i,l,t]$, for the rest of the paper.} 
\begin{equation} \label{equ:network_it}
	\mathcal{N}[i,t] = \mathcal{M}\bigl[ i,l[i,t],t \bigr].
\end{equation}
An example of $\mathcal{N}[i,t]$ is given in Fig.~\ref{fig:resourceblock}.

\subsection{Network-Time Route as Action} \label{sec:networktime}

  After describing a user's network availability and mobility, we define his \emph{action} as his network selections across multiple time slots, which is referred to as \emph{network-time route} define below. 
  Let $\tilde{\mathcal{R}}_i$ be the set of all possible network-time routes of user $i$.
	Given user $i$'s mobility pattern $\boldsymbol{\theta}_i$,  we let $\mathcal{R}_i(\boldsymbol{\theta}_i) \subseteq \tilde{\mathcal{R}}_i$ be the set of all \emph{feasible} network-time route of user $i$ define as follows.

\begin{definition}[Feasible Network-Time Route]  \label{def:networktimeroute}
  Given user $i$'s mobility pattern $\boldsymbol{\theta}_i$, his \emph{feasible network-time route} is a sequence
\begin{equation} \label{equ:action}
  \boldsymbol{r}_i = \Bigl((n_i^{1},t_i^{1}),(n_i^2,t_i^2),\ldots,(n_i^{Q_i},t_i^{Q_i})\Bigr) \in \tilde{\mathcal{R}}_i,
\end{equation}
which indicates user $i$'s network selections in all time slots, except those time slots when user $i$ is in the middle of network switching and is not associated with any network.
It satisfies the following conditions: 
 \begin{enumerate}
  \item Causality: $1=t_i^{1}<t_i^{2}<\ldots<t_i^{Q_i} \leq T$.
	
  \item Eligibility: $n_i^{q} \in \mathcal{N}[i,t_i^q]$, for each $q \in \{ 1, \ldots, Q_i \}$.
	
  \item Switching time: $t_i^{q+1} - t_i^q = \delta[n_i^q, n_i^{q+1}] + 1$, for each $q \in \{ 1, \ldots, Q_i-1 \}$.
 \end{enumerate}
\end{definition}

  Condition $1)$ accounts for the fact that time is always increasing. 
	Condition $2)$ ensures that user $i$ is eligible to select the networks according to their availabilities as defined in \eqref{equ:network_it}.
	Condition $3)$ ensures that the time difference between successive elements in the sequence of network-time route is consistent with the switching time between the corresponding networks. 
	More specifically, when $n_i^q = n_i^{q+1}$, it means that user $i$ keeps using the same network at the next time slot. Without involving any network switching, we have $t_i^{q+1}-t_i^q = 1$ since $\delta[n,n] = 0, \fa{n}{N}$. 
	When $n_i^q \neq n_i^{q+1}$, it means that user $i$ switches from network $n_i^q$ to $n_i^{q+1}$. Therefore, user $i$ can use network $n_i^{q+1}$ after finishing the switching process, which takes switching time of $\delta[n_i^q,n_i^{q+1}]$.

\begin{figure}[t]
 \centering
	 \includegraphics[width=6cm, trim = 4.5cm 2.2cm 4cm 1cm, clip = true]{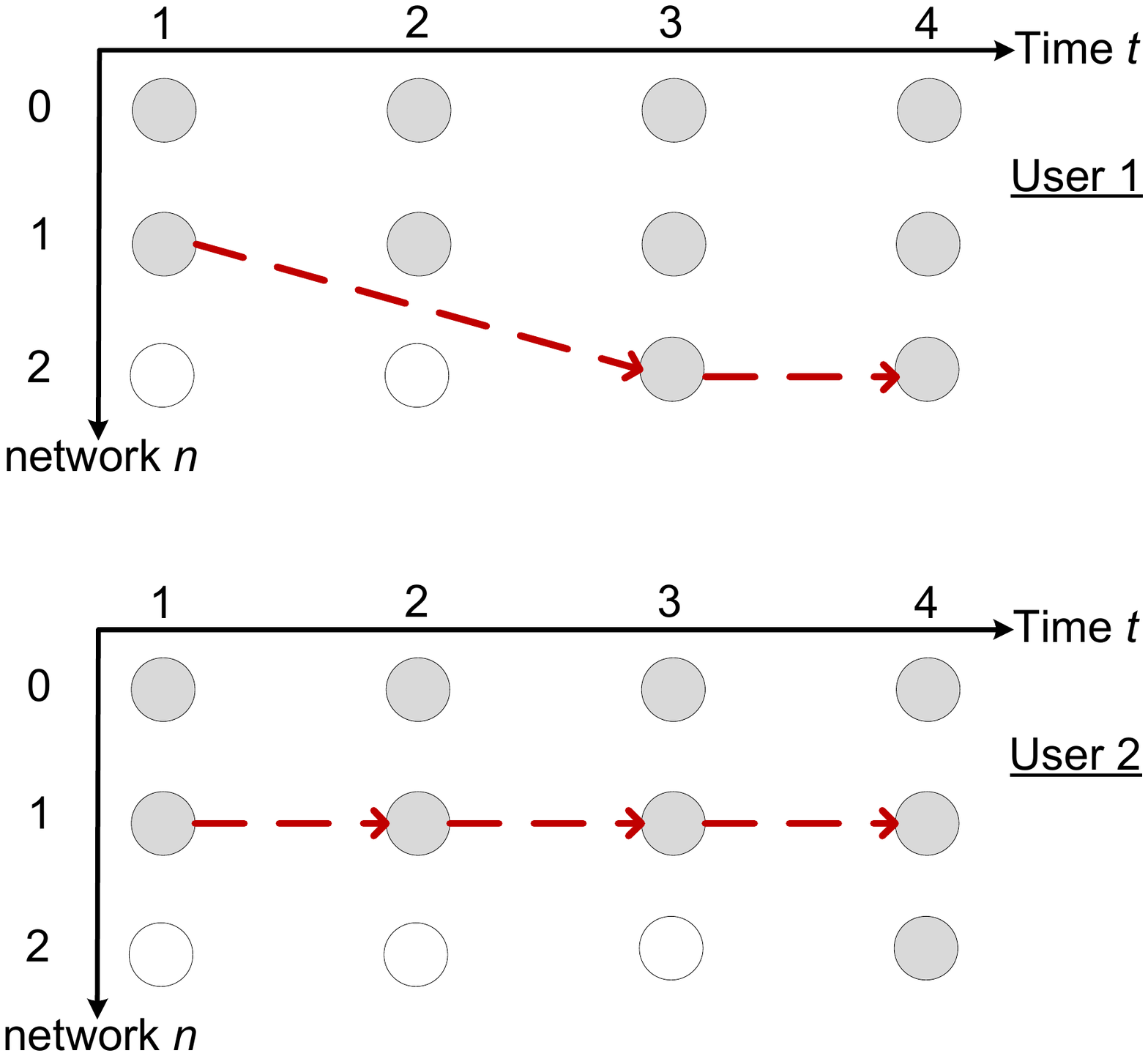} 
 \caption{The network-time routes chosen by the two users when switching time $\delta[1,2] = 1$. } 
\label{fig:route}
\end{figure}

  To facilitate the introduction of the user's utility function in the next subsection, 
  we define the network-time points of a feasible network-time route as the network-time selections along it.
	
\begin{definition}[Network-time points]
  Given a feasible route $\boldsymbol{r}_i \in \mathcal{R}_i(\boldsymbol{\theta}_i)$ in \eqref{equ:action}, we define its network-time points as the set
\begin{equation} \label{equ:vertices}
	\mathcal{V}(\boldsymbol{r}_i) = \Bigl\{(n_i^{1},t_i^{1}),(n_i^2,t_i^2),\ldots,(n_i^{Q_i},t_i^{Q_i}) \Bigr\}. 
\end{equation}
  The set can also be represented as the $Q_i-1$ network-time point pairs 
%
\begin{equation} \label{equ:edges}
	\mathcal{E}(\boldsymbol{r}_i) = \Bigl\{ \bigl( (n_i^{q},t_i^{q}),(n_i^{q+1},t_i^{q+1}) \bigr): q = 1,\ldots,Q_i-1 \Bigr\},
\end{equation}
which are the consecutive pairs of network-time points visited by user $i$ in route $\bs{r}_i$.
\end{definition}

\begin{figure*}[t]
\centering
\begin{minipage}[t]{0.45\linewidth}
       \includegraphics[width=7cm, trim = 3.5cm 5cm 5cm 5cm, clip = true]{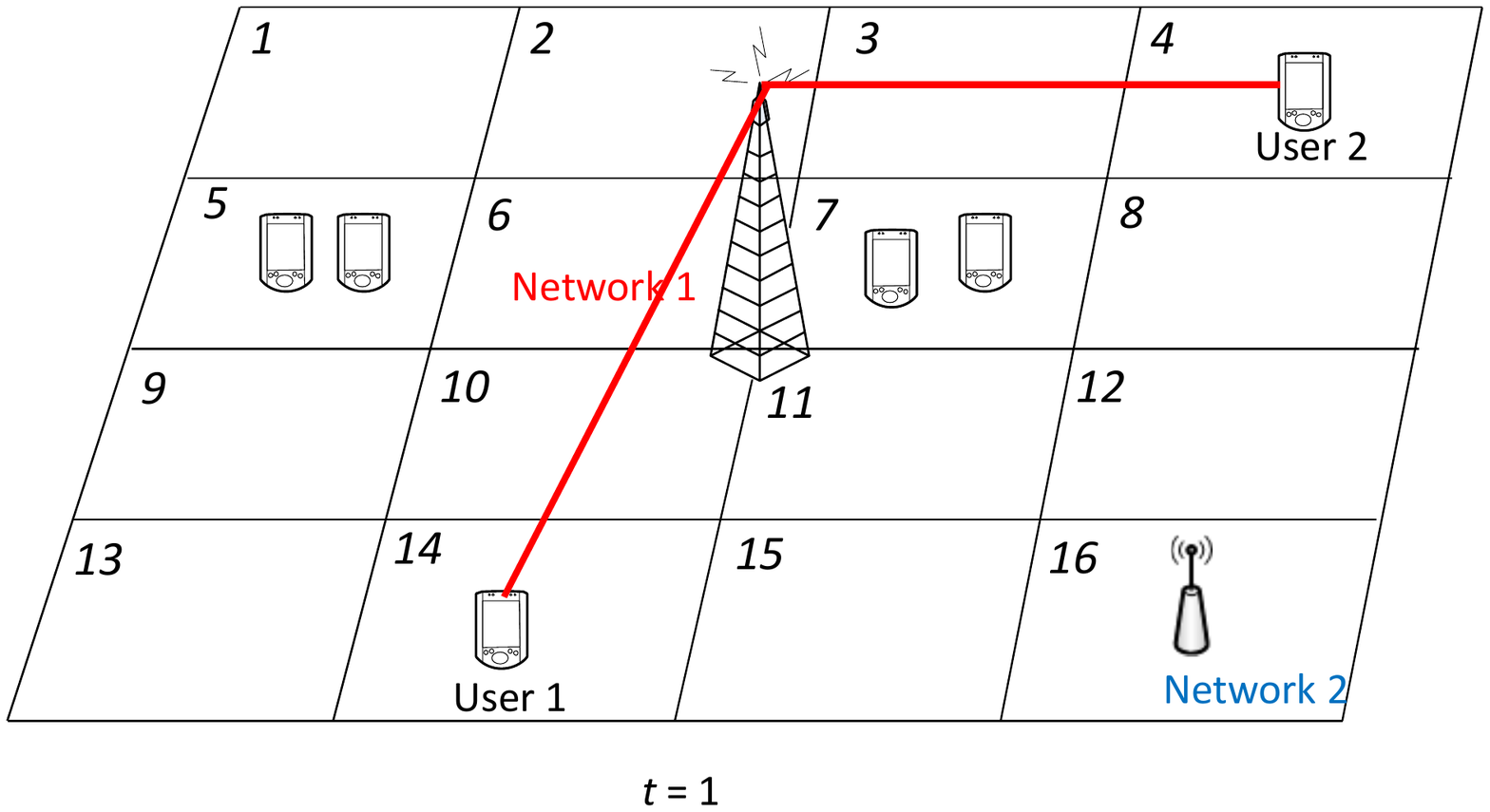}
\end{minipage}
\quad
\begin{minipage}[t]{0.45\linewidth}
       \includegraphics[width=7cm, trim = 3.5cm 5cm 5cm 5cm, clip = true]{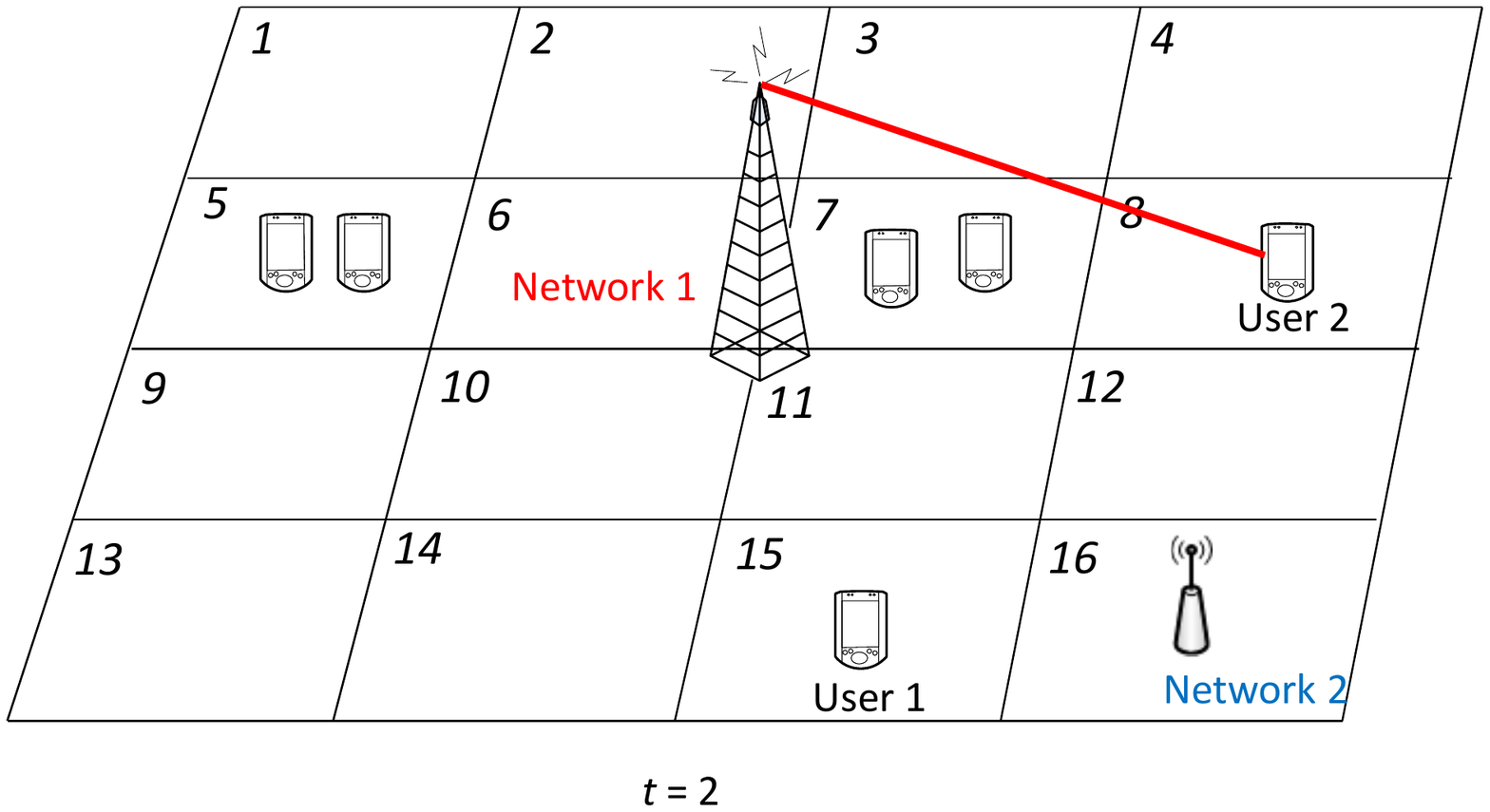}
\end{minipage}
\newline
\begin{minipage}[t]{0.45\linewidth}
       \includegraphics[width=7cm, trim = 3.5cm 5cm 5cm 5cm, clip = true]{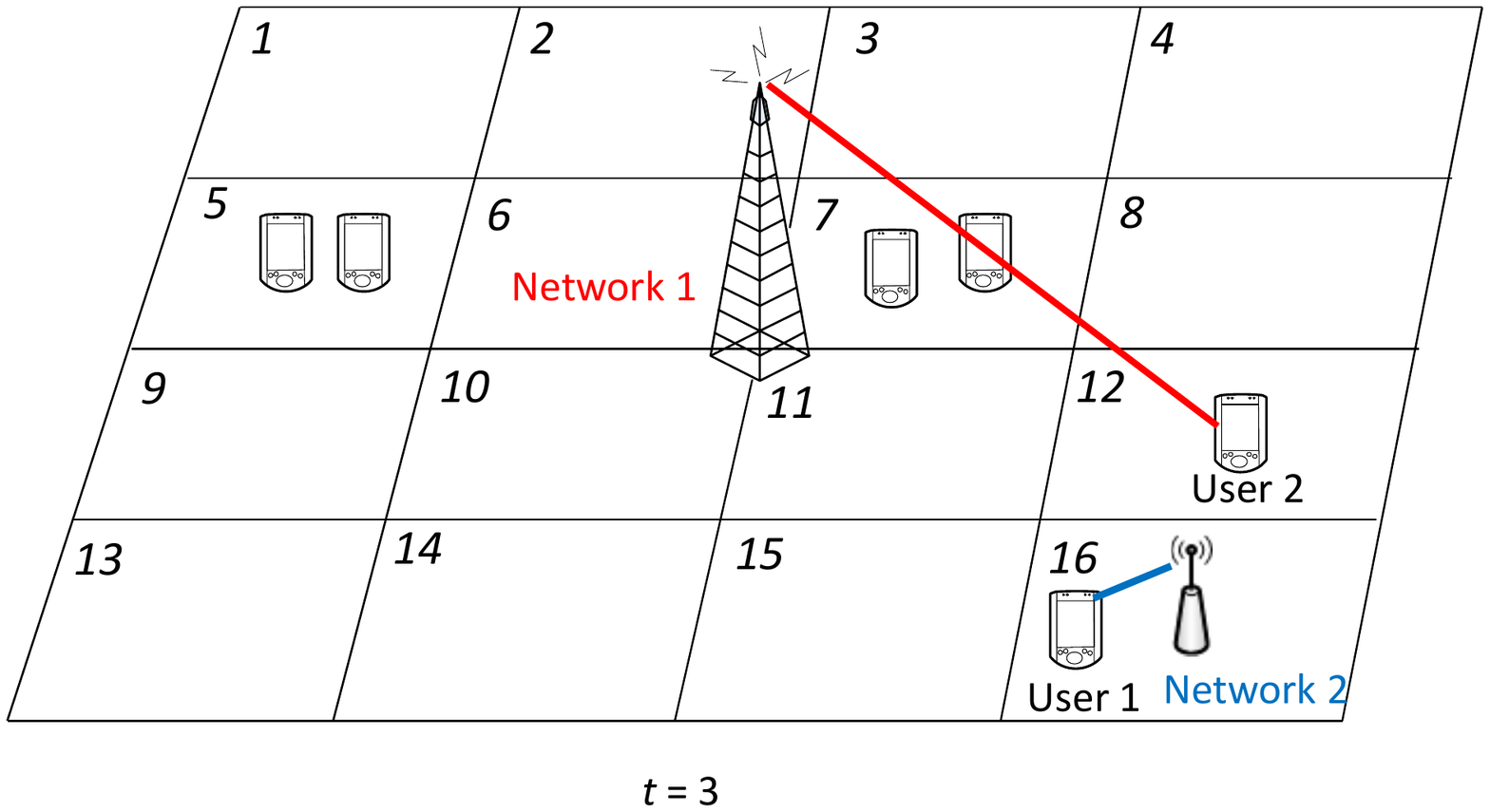}
\end{minipage}	
\quad
\begin{minipage}[t]{0.45\linewidth}
       \includegraphics[width=7cm, trim = 3.5cm 5cm 5cm 5cm, clip = true]{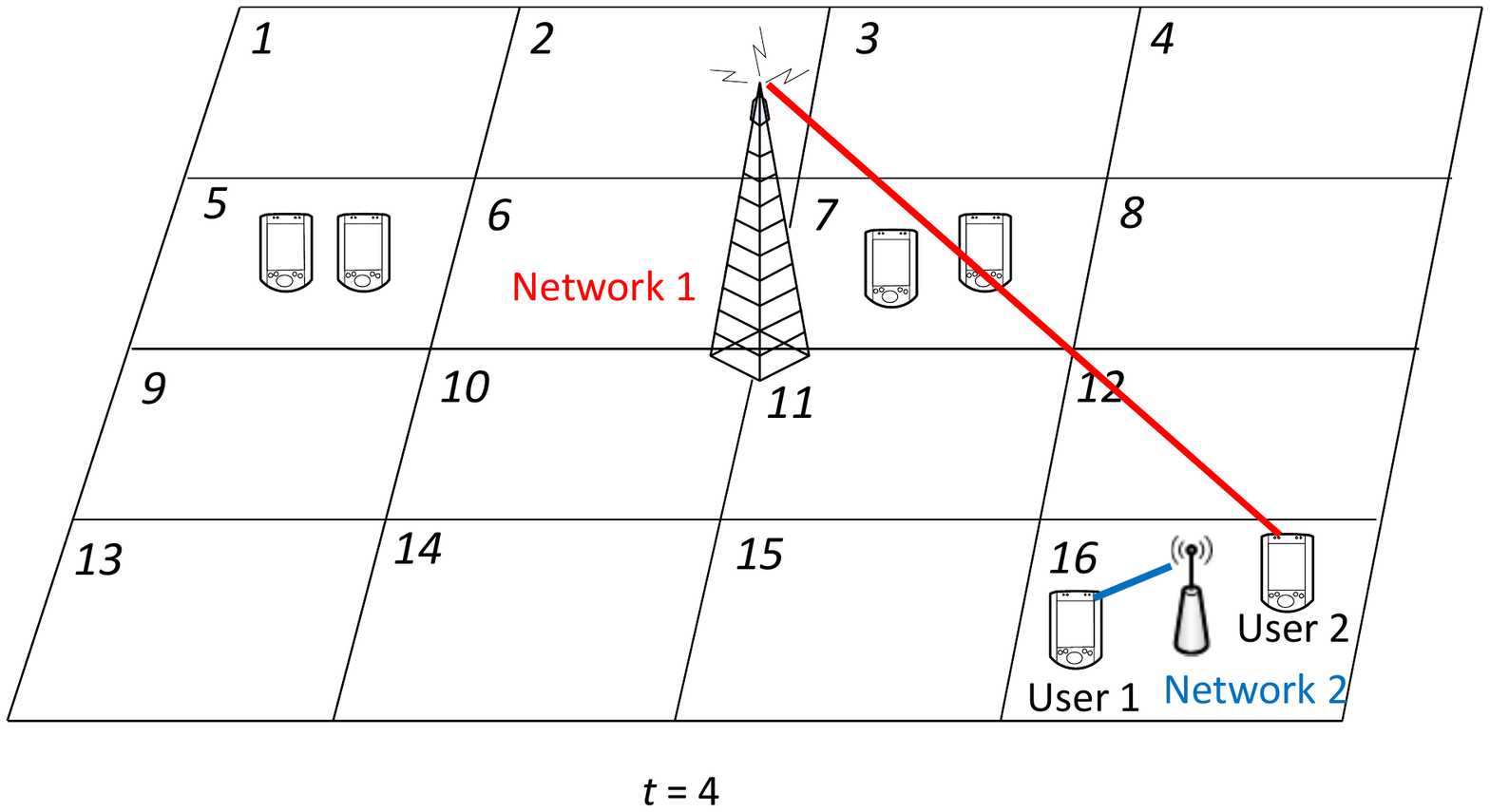}
\end{minipage}
\vspace{-0.2cm}
\caption{The network selections of users $1$ and $2$ at different locations in the four time slots.}
\label{fig:spatial}
\end{figure*}

  An example of the feasible network-time routes is shown in Fig.~\ref{fig:route}. In this example, we have $\boldsymbol{r}_1 = \bigl( (1,1), (2,3), (2,4) \bigr)$, so $\mathcal{V}(\boldsymbol{r}_1) = \{(1,1), (2,3), (2,4)\}$ and $\mathcal{E}(\boldsymbol{r}_1) = \Bigl\{\bigl( (1,1), (2,3) \bigr), \bigl( (2,3), (2,4) \bigr) \Bigr\}$.
	The pair of network-time points $\bigl( (1,1), (2,3) \bigr)$ means that user $1$ accesses network $1$ at time slot $1$, and switches to  network $2$ at time slot $3$ after taking one time slot of switching time. 
	The pair of network-time points $\bigl( (2,3), (2,4) \bigr)$ denotes that user $1$ accesses network $2$ at time slot $3$, and keeps using network $2$ at time slot $4$.   
	The corresponding network selections of users at $1$ and $2$ at different locations and time slots are illustrated in Fig.~\ref{fig:spatial}.
	

\subsection{Utility Function} \label{sec:utility}

  For the design of a practical congestion-aware network selection algorithm, a user's utility function should take both the network congestion and the negative impact of ping-pong effect into account.
	Let $\omega[(n, t), \boldsymbol{r}, \boldsymbol{\theta}] = |j \in \mathcal{I}: (n,t) \in \mathcal{V}(\boldsymbol{r}_j), \boldsymbol{r}_j \in \mathcal{R}_j(\boldsymbol{\theta}_j)|$ be the \emph{network congestion level}, which counts the number of network-time routes chosen in action profile $\boldsymbol{r} = (\boldsymbol{r}_1,\dots,\boldsymbol{r}_I)$ that pass through the network-time point $(n, t)$  under mobility patterns $\boldsymbol{\theta} = (\boldsymbol{\theta}_1,\dots,\boldsymbol{\theta}_I)$. 
	In other words, it denotes the total number of users accessing network $n$ at time $t$.
	For a route $\boldsymbol{r}_i \in \mathcal{R}_i(\boldsymbol{\theta}_i)$ in \eqref{equ:action}, user $i$'s utility function consists of two parts:

\begin{itemize}
	\item Total throughput: The summation of user $i$'s achieved throughput over all the network-time points along route $\bs{r}_i$ (i.e., set $\mathcal{V}(\boldsymbol{r}_i)$ in \eqref{equ:vertices}). At each network-time point $(n, t)$, the achieved throughput of user $i$ is the network capacity $\mu[n]$ divided by the network congestion level $\omega[(n, t), \boldsymbol{r}, \boldsymbol{\theta}]$. 
	
	\item Total switching cost: The summation of the switching costs for network switching of every network-time point pairs in route $\boldsymbol{r}_i$ (i.e., set $\mathcal{E}(\boldsymbol{r}_i)$ in \eqref{equ:edges}). 
	More specifically, for each network-time point pair $\boldsymbol{e} = \bigl( (n_i^{q},t_i^{q}),(n_i^{q+1},t_i^{q+1}) \bigr) \in \mathcal{E}(\boldsymbol{r}_i)$, we define its switching cost as
\begin{equation}
	g[\boldsymbol{e}] = g \bigl[ (n_i^{q},t_i^{q}),(n_i^{q+1},t_i^{q+1}) \bigr] = c[n_i^{q},n_i^{q+1}].
\end{equation}
\end{itemize}	

  Overall, given the action profile $\boldsymbol{r}$ and mobility patterns $\boldsymbol{\theta}$ of all users, if $\boldsymbol{r}_i \in \mathcal{R}_i(\boldsymbol{\theta}_i)$, user $i$'s \emph{utility function} is 
\begin{equation} \label{equ:utility_route}
	U_i(\boldsymbol{r},\boldsymbol{\theta}) = \displaystyle\sum_{(n,t) \in \mc{V}(\boldsymbol{r}_i)} \frac{\mu[n]}{\omega[(n, t), \boldsymbol{r}, \boldsymbol{\theta}]} - \sum_{\boldsymbol{e} \in \mathcal{E}(\boldsymbol{r}_i)} g[\boldsymbol{e}].
\end{equation}
\section{Centralized Network Allocation} \label{sec:centralized}

  One natural formulation is to consider the centralized network allocation that maximizes the users' aggregate utilities. 
	However, we show that this would be an NP-hard problem even in the special case of deterministic mobility pattern, where the users' mobility patterns are known.\footnote{It should be noted that in the general case with random mobility patterns, we will consider the maximization of the users' aggregate expected utilities, which involves solving a number of network allocation problems under the deterministic mobility pattern in the form of problem \eqref{equ:socialopt}. In other words, if problem \eqref{equ:socialopt} is NP-hard, then the problem under the random mobility case is also NP-hard.} 


  We first formally define a centralized socially optimal network allocation as follows.

\begin{definition}[Socially Optimal Network Allocation] 
  Given the deterministic mobility patterns $\boldsymbol{\theta}$, an action profile $\boldsymbol{r}^*$ is a socially optimal network allocation if it maximizes the social welfare: 
\begin{equation} \label{equ:socialopt}
	\boldsymbol{r}^* = \argmax_{\boldsymbol{r}_i \in \mc{R}_i(\bs{\theta}_i), \, \forall \, i \in \mathcal{I}} \mathbb{W}(\boldsymbol{r},\boldsymbol{\theta}) \triangleq \sum_{i \in \mc{I}} U_i(\boldsymbol{r},\boldsymbol{\theta}), 
\end{equation}
\end{definition}
where the social welfare $\mathbb{W}(\boldsymbol{r},\boldsymbol{\theta})$ is defined as the users' aggregate utility.

  With Assumption \ref{ass:idle} in Section \ref{sec:network}, we can show that there always exists a socially optimal network allocation that each network-time point is chosen by at most one user.

\begin{lemma} \label{lem:socialwelware}
  Under Assumption \ref{ass:idle} and the given deterministic mobility patterns $\boldsymbol{\theta}$, there always exists a socially optimal network allocation $\boldsymbol{r}^*$ such that the congestion level $\omega[(n, t), \boldsymbol{r}^*, \boldsymbol{\theta}] \leq 1$, for all network-time points $(n,t) \in \mathcal{N} \times \mc{T}$.	
\end{lemma}

	The proof of Lemma \ref{lem:socialwelware} is given in Appendix \ref{app:socialwelware}.
	With Lemma \ref{lem:socialwelware}, we can show that the problem of finding a socially optimal network allocation is NP-hard.

\begin{theorem} \label{thm:np_hard}
  The problem of finding a socially optimal network allocation in \eqref{equ:socialopt} is NP-hard.
\end{theorem}

	The proof of Theorem \ref{thm:np_hard} is given in Appendix \ref{app:np_hard}.
  Thus, solving the centralized network allocation problem is \emph{infeasible} for practical system with a potentially large number of users, networks, and mobility patterns.
	Moreover, it is more practical to study the scenario that the users are autonomously selecting the networks themselves, rather than the operator controlling their network choices.
	This motivates us to formulate the \emph{distributed} network selection problem as a non-cooperative game, as we discuss next. 

\section{Distributed Network Selection Game} \label{sec:game}

  In this section, we formulate the users' network selections with incomplete information as a distributed network selection game (NSG).
	We first describe the non-cooperative game formulation in Section \ref{sec:game_incompleteinfo}.
	We then show that it is a Bayesian potential game with the finite improvement property and derive its exact potential function in closed-form in Section \ref{sec:game_bayesian}.
	Finally, we proposed a distributed network selection (DNS) algorithm to coordinate the users' decisions in Section \ref{sec:algo}.

\subsection{Network Selection Game with Incomplete Information} \label{sec:game_incompleteinfo}

  In practice, each user may only have \emph{incomplete} information of the types (i.e., mobility patterns) of all the users (even including himself) at the beginning of the time period. 
	More specifically, we assume that the users' types $\bs{\theta}$ follow a known \emph{prior} probability distribution $p(\boldsymbol{\theta}_1,\dots,\boldsymbol{\theta}_I)$.\footnote{It is possible that this prior information on mobility patterns can be obtained from the mobile operator. However, as we will discuss in Section \ref{sec:algo}, the actual implementation of the DNS algorithm (i.e., Algorithm \ref{algo:ns_mu}) only requires the aggregate network usage statistics from the operator, instead of the detailed users' mobility information, so there are no privacy concerns in the proposed algorithm.} 
	In addition, the utility functions, available actions, possible types, and the prior distributions of user types are assumed to be public information.
	We then formulate the network selection game as a \emph{Bayesian game} \cite{shoham_ma08} as follows.

\begin{definition}[Network Selection Game] 
  A network selection game is a tuple $\Omega = (\mathcal{I}, \tilde{\mathcal{R}}, \boldsymbol{\Theta}, p, \boldsymbol{U})$ defined by:

\begin{itemize}
	\item Players: The set of users $\mathcal{I}$.
	
	\item Actions: The set of action profiles of all the users is $\tilde{\mathcal{R}} = \tilde{\mathcal{R}}_1 \times \ldots \times \tilde{\mathcal{R}}_I$. 
	
	\item Types: The set of type space of all the users is $\boldsymbol{\Theta} = \boldsymbol{\Theta}_1 \times \ldots \times \boldsymbol{\Theta}_I$, where $\boldsymbol{\Theta}_i$ is the set of possible mobility patterns of user $i$.
	
	\item Prior information: The common prior probability $p(\boldsymbol{\theta}_1,\dots,\boldsymbol{\theta}_I)$ over the types of all users.  
	
	\item Utilities: The vector $\boldsymbol{U} = (U_i, \, \forall \, i \in \mathcal{I})$ contains the utility functions of the users defined in \eqref{equ:utility_route}.
	
\end{itemize}
\end{definition}

  Under this Bayesian game setting, user $i$'s \emph{strategy} is a mapping $s_i: \Theta_i \rightarrow \tilde{\mathcal{R}}_i$, which specifies user $i$'s action for each possible type. 
	In other words, $\bs{s}_i(\bs{\theta}_i)$ specifies the network-time route that user $i$ should choose given his mobility pattern $\bs{\theta}_i$.
	Let $\boldsymbol{\theta} = (\boldsymbol{\theta}_1,\dots,\boldsymbol{\theta}_I) \in \boldsymbol{\Theta}$ be the users' type profile.
	So $\bs{s}(\bs{\theta}) = \bigl( \bs{s}_1(\bs{\theta}_1), \ldots, \bs{s}_I(\bs{\theta}_I) \bigr)$ is the action profile of all the users given that their type profile is $\bs{\theta}$.
	The \emph{expected utility} of user $i$ under strategy profile $\boldsymbol{s}$ is
\begin{equation} \label{equ:expectedutility} 
	EU_i(\boldsymbol{s}) = \sum_{\boldsymbol{\theta} \in \Theta} U_i \bigl( \boldsymbol{s}(\boldsymbol{\theta}), \boldsymbol{\theta} \bigr) \, p(\boldsymbol{\theta}).
\end{equation}

  Let $\boldsymbol{s}_{-i} = (\boldsymbol{s}_1, \ldots, \boldsymbol{s}_{i-1}, \boldsymbol{s}_{i+1}, \ldots, \boldsymbol{s}_I)$ denote the strategies of all the users except user $i$. A strategy profile can be written as $\boldsymbol{s} = (\boldsymbol{s}_i, \boldsymbol{s}_{-i})$.
  Let $\mathcal{S}_i$ be the strategy space of user $i$ and let $\mathcal{S} = \mathcal{S}_1 \times \ldots \times \mathcal{S}_I$ be the strategy space of all the users.
  We define the Bayesian Nash equilibrium \cite{shoham_ma08} as follows.
	
\begin{definition}[Bayesian Nash equilibrium]	 \label{def:bayesianne}
  The strategy profile $\boldsymbol{s}^*$ is a pure strategy Bayesian Nash equilibrium (BNE)\footnote{Alternatively, we may define the BNE based on each user $i$'s \emph{ex-interim} expected utility $EU_i(\bs{s}_i, \bs{s}_{-i}^*|\bs{\theta}_i)$ \cite{shoham_ma08}, where user $i$ knows his own type $\bs{\theta}_i$. In this case, the strategy profile $\boldsymbol{s}^*$ is a pure strategy BNE if $\bs{s}_i^* = \argmax_{\bs{s}_i \in \mc{S}_i} EU_i(\bs{s}_i, \bs{s}_{-i}^*|\bs{\theta}_i), \, \forall \, \bs{\theta}_i \in \Theta_i, i \in \mathcal{I}$. However, Theorem I in \cite{harsanyi_gw68} showed that such an alternative definition is equivalent to the BNE defined in \eqref{equ:bne_exante}, where user $i$ knows nothing about any user's actual type (including his own type).} if
\begin{equation} \label{equ:bne_exante}
	\bs{s}_i^* = \argmax_{\bs{s}_i \in \mc{S}_i} EU_i(\bs{s}_i, \bs{s}_{-i}^*), \, \forall \, i \in \mathcal{I}.	
\end{equation}
\end{definition}

\subsection{Bayesian Potential Game} \label{sec:game_bayesian}

  In general, it is difficult to establish the analytical results of the BNE. 
  Nevertheless, in this subsection, we are able to show that $\Omega$ is a \emph{Bayesian potential game} \cite{facchini_cm97}, which exhibits the finite improvement property. It thus implies the existence of and the convergence to the BNE.
	
	First, we present the definition of a Bayesian potential game \cite{facchini_cm97}.

\begin{definition}[Bayesian Potential Game]
  Bayesian Game $\Omega$ is a \emph{Bayesian potential game} if there exists an exact potential function $\Phi(\boldsymbol{r}, \boldsymbol{\theta})$ such that
\begin{equation} \label{equ:potential_deterministic_def}
\begin{split}
	U_i(\boldsymbol{r}_i,\! \boldsymbol{r}_{-i}, \boldsymbol{\theta}) \! - \! U_i(\boldsymbol{r}_i',\! \boldsymbol{r}_{-i}, \boldsymbol{\theta}) 
	 =  \Phi(\boldsymbol{r}_i, \boldsymbol{r}_{-i}, \boldsymbol{\theta}) \! - \! \Phi(\boldsymbol{r}_i', \boldsymbol{r}_{-i}, \boldsymbol{\theta}), \\
	\forall \, \boldsymbol{r}_i, \boldsymbol{r}_i' \in \mathcal{R}_i(\boldsymbol{\theta}_i), \boldsymbol{\theta} \in \Theta, i \in \mathcal{I}.\footnote{We assume that the users are rational, who aim to maximize their expected utility. As a result, we can focus our attention on the set of feasible routes $\mathcal{R}_i(\boldsymbol{\theta}_i)$ of each user $i$.}
\end{split}	
\end{equation}
\end{definition}
	
  In other words, Bayesian potential game is a Bayesian game, where the change in the value of the utility function is equal to the change in the value of the potential function when the action profile changes.

\begin{theorem} \label{thm:bpg}
  Game $\Omega$ is a Bayesian potential game with the exact potential function given by
\begin{equation} \label{equ:potential_deterministic}
	\Phi(\boldsymbol{r}, \boldsymbol{\theta}) = \sum_{(n,t) \in \mathcal{N} \times \mathcal{T}} \sum_{q=1}^{\omega[(n,t), \boldsymbol{r}, \boldsymbol{\theta}]} \frac{\mu[n]}{q} - \sum_{i \in \mathcal{I}} \sum_{\boldsymbol{e} \in \mathcal{E}(\boldsymbol{r}_i)} g[\boldsymbol{e}].
\end{equation}
\end{theorem}

  The proof of Theorem \ref{thm:bpg} is given in Appendix \ref{app:bpg}.

\subsubsection{Properties of the Bayesian Potential Game}

 Before stating the convergence properties of the users' interactions, let us recall some definitions \cite{vocking_cg06}.

\begin{definition}[Better and best response updates]
   Starting from a strategy profile $\boldsymbol{s} = (\boldsymbol{s}_i, \boldsymbol{s}_{-i})$, a \emph{better response update} is an event where a single user $i$ changes his strategy from $\boldsymbol{s}_i \in \mathcal{S}_i$ to $\boldsymbol{s}_i' \in \mathcal{S}_i$, and increases his expected utility as a result, i.e., $EU_i(\boldsymbol{s}_i', \boldsymbol{s}_{-i}) > EU_i(\boldsymbol{s}_i, \boldsymbol{s}_{-i})$.

   A \emph{best response update} is a special type of better response update, where the newly selected strategy $\boldsymbol{s}_i'$ maximizes user $i$'s expected utility among user $i$'s all possible better response updates. 
\end{definition}

\begin{definition}[Finite improvement property] \label{def:FIP}  
   A Bayesian game possesses the \emph{finite improvement property} (FIP) when asynchronous\footnote{Asynchronous updates imply that there will be no two users updating their strategies at the same time.} better response updates always converge to a BNE (defined in Definition \ref{def:bayesianne}) within a finite number of steps, irrespective of the initial strategy profile or the users' updating order.
\end{definition}

 The FIP implies that better response updating always leads to a pure strategy BNE, which implies the existence of pure strategy BNE.
 As a result, it ensures the efficient convergence of the users' network selection to an equilibrium point and thus the stability of the system.

\begin{theorem} \label{thm:fip}
  Game $\Omega$ possesses the finite improvement property.
\end{theorem}

  The proof of Theorem \ref{thm:fip} is given in Appendix \ref{app:fip}.

\subsection{Algorithm Design} \label{sec:algo}

  With the nice FIP just derived, we propose a distributed network selection (DNS) algorithm (i.e., Algorithm \ref{algo:ns_mu}) for the users to make their network selection decisions autonomously.
	This algorithm relies on the network information obtained from the operator in Algorithm \ref{algo:ns_mo}.
	Different from the Bayesian game setting discussed above that requires the prior information of the users' mobility, we design the DNS algorithm in a way that it only requires the users to report their network selection statistics. Thus, it eliminates the privacy concern of asking users to reveal their mobility information. 

\subsubsection{Algorithm \ref{algo:ns_mu}}	
	
	To initialize, a user inputs his destination on the mobile device (line 1). Based on his current location (e.g., obtained from global positioning system (GPS)) and mobility history, the device can compute the user's possible trajectories and the corresponding probabilities (line 2). The device then queries the subscription plan (line 3) and network parameters (line 4) on behalf of the users to determine the available network resources. 
	
	Next, we describe the best response update of user $i$:


\begin{algorithm} [t] \small
\caption{\emph{Distributed network selection (DNS) algorithm for user $i \in \mathcal{I}$.}}
\begin{algorithmic} [1] 

\STATEx \underline{Initialization}

\STATE User's input: Destination. 

\STATE Compute the set $\Theta_i$ of possible trajectories and the corresponding probability $p(\boldsymbol{\theta}_i) \geq 0$ for each $\boldsymbol{\theta}_i \in \Theta_i$ based on the user's current location and mobility history, such that $\sum_{\bs{\theta_i} \in \Theta_i} p(\boldsymbol{\theta}_i) = 1$.

\STATE Query user $i$'s subscription plan $\mathcal{M}[i,l,t]$ $\forall \, l \in \mathcal{L}, t \in \mathcal{T}$ from the operator's database. 

\STATE Query network parameters from the operator's database: network capacity $\mu[n], \, \forall \, n \in \mathcal{N}$,  switching time $\delta[n,n']$, and switching cost $c[n,n'], \, \forall \, n,n' \in \mathcal{N}, n \neq n'$.

\STATEx \underline{Planning Phase: Bayesian Network Selection Game}

\STATE \textbf{repeat}


\STATE $ \ \ \ $ \textbf{if} $\tau \in \Gamma_i$

\STATE $ \ \ \ \ \ $ \textbf{for} $\bs{\theta_i} \in \Theta_i$

\STATE $ \ \ \ \ \ \ \ $ Compute network availabilities $\mathcal{N}[i,t] := \mathcal{M}\bigl[ i,l[i,t],t \bigr],$ \STATEx \quad\quad\quad\quad $\, \forall \, t \in \mathcal{T}$ for trajectory $\boldsymbol{\theta}_i = (l[i,t], \forall \, t \in \mc{T})$. 

\STATE $ \ \ \ \ \ \ \ $ Determine set $\mathcal{R}_i(\boldsymbol{\theta}_i)$ of feasible network-time routes \STATEx \quad\quad\quad\quad from $\mathcal{N}[i,t]$ and switching time $\delta[n,n'], \forall \, n, n' \in \mc{N}$ \STATEx \quad\quad\quad\quad by Definition \ref{def:networktimeroute}.

\STATE $ \ \ \ \ \ \ \ $ Obtain pmf $p_{-i}[q,(n,t)]$ for $q = 0, \ldots, I-1$ for all \STATEx \quad\quad\quad\quad $(n,t) \in \mathcal{N} \times \mathcal{T}$ from the operator (see Algorithm \ref{algo:ns_mo}). 

\STATE $ \ \ \ \ \ \ \ $ Perform a best response update for type $\bs{\theta_i}$, by \STATEx \quad\quad\quad\quad identifying a route $\bs{r}_i \in \mathcal{R}_i(\boldsymbol{\theta}_i)$ that maximizes \STATEx \quad\quad\quad\quad user $i$'s expected utility given that his type is $\bs{\theta_i}$: 
\begin{equation} \label{equ:utility_br}
	u_i(\boldsymbol{r}_i|\boldsymbol{\theta_i}) := \hspace{-0.3cm} \displaystyle\sum_{(n,t) \in \mc{V}(\boldsymbol{r}_i)} \sum_{q=0}^{I-1} \left( \frac{\mu[n]}{q + 1} \right) p_{-i}[q,(n,t)] - \sum_{\boldsymbol{e} \in \mathcal{E}(\boldsymbol{r}_i)} g[\boldsymbol{e}]. 
\end{equation}

\STATE $ \ \ \ \ \ \ \ $ Update the strategy $\boldsymbol{s}_i(\boldsymbol{\theta_i}) := \bs{r}_i$.

\STATE $ \ \ \ \ \ $ \textbf{end for}

\STATE $ \ \ \ \ \ $ Compute the network selection statistics under strategy $\boldsymbol{s}_i$:
\begin{equation} \label{equ:pint}
	p_i[n,t] := \sum_{\bs{\theta_i} \in \Theta_i} p(\bs{\theta_i}) \times \textbf{1}_{(n,t) \in \mc{V}(\bs{s}_i(\bs{\theta_i}))}, \, \forall \, (n,t) \in \mc{N} \times \mc{T},
\end{equation}
\quad\quad\quad where $\textbf{1}_{\{.\}}$ is the indicator function.

\STATE $ \ \ \ \ \ $ Report the network selection statistics \STATEx \quad\quad\quad\quad $\bs{p}_i := (p_i[n,t], \, \forall \, (n,t) \in \mathcal{N} \times \mathcal{T})$ to the operator.

\STATE $ \ \ \ $ \textbf{end if}

\STATE \textbf{until} $\tau \geq \tau^{\text{max}}$.

\STATEx \underline{Network Selection Phase}

\STATE User $i$ determines his actual trajectory $\boldsymbol{\theta}_i$.

\STATE Select networks in different time slots based on action $\boldsymbol{s}_i(\boldsymbol{\theta_i})$. 

\end{algorithmic} \label{algo:ns_mu}
\end{algorithm}

\begin{itemize}
	\item \emph{Network information collection}:
	Let $\Gamma_i$ be the set of time slots (line 6) that user $i$ updates his strategy, which corresponds to a network-time route for each of his possible type $\bs{\theta}_i$ (line 7).
	Based on the user, location, and time availability of the networks, the device computes user $i$'s available networks (line 8) and his set of feasible network-time routes (line 9).\footnote{The operator can obtain the Wi-Fi network information if the Wi-Fi APs are deployed by the operator. For other Wi-Fi APs, it is possible to obtain this information from other operators under roaming agreements.}
	
	\item \emph{Best response computation}: Let $p_{-i}[q,(n,t)]$ be the probability that the network-time point $(n,t)$ would be occupied by $q$ users (i.e., the congestion level is $q$) given a set $\mc{I} \backslash \{i\}$ of users, where $q = 0, \ldots, I-1$.
	With this information on the probability mass function (pmf) of congestion level in each network at different time from the operator in Algorithm \ref{algo:ns_mo} (line 10), each user performs a best response update to maximize his expected utility\footnote{For notational simplicity, we define $u_i(\boldsymbol{r}_i|\boldsymbol{\theta_i})$ as the expected utility in the algorithm, which is actually equal to $EU_i(\bs{s}_i, \bs{s}_{-i}|\bs{\theta}_i)$.} $u_i(\boldsymbol{r}_i|\boldsymbol{\theta_i})$ for choosing route $\bs{r}_i$ given that his type is $\bs{\theta_i}$ (line 11), where the first and second terms on the right hand side of \eqref{equ:utility_br} represent the expected throughput and switching cost for choosing route $\bs{r}_i$, respectively.
  Each best response update can be computed by applying a shortest path algorithm \cite{weiss_ds97}, such as Bellman-Ford algorithm, and can be performed in an asynchronous fashion by the users until reaching the iteration limit\footnote{According to FIP, the algorithm will always converge in a finite number of iterations. However, we add this iteration limit to ensure that the convergence does not take too long, thus allows the tradeoff between the performance and convergence time. In our simulation in Section \ref{sec:pe}, we choose $\tau^{\text{max}} = 20$.} of $\tau^{\text{max}}$, which represents the maximum number of iterations that the algorithm will run (line 17).
	
	\item \emph{Information exchange}: Each user needs to report his individual network selection statistics to the operator, so that the operator can calculate the aggregate network congestion statistics by Algorithm \ref{algo:ns_mo}.
	Let $p_i[n,t]$ be the probability that user $i$ would access network-time point $(n,t)$ under strategy $\bs{s}_i$. 
	After the strategy $\bs{s}_i$ is determined, the device computes $p_i[n,t]$ by summing the probability $p(\bs{\theta}_i)$ that network-time point $(n,t)$ is chosen in the action $\bs{s}_i(\bs{\theta}_i)$ in \eqref{equ:pint} (line 14) and reports the network selection statistics $\boldsymbol{p}_i = (p_i[n,t], \, \forall \, (n,t) \in \mathcal{N} \times \mathcal{T})$ of all the network-time points to the operator (line 15).
	
	\item \emph{Network selection}: After that, once user $i$' actual trajectory $\boldsymbol{\theta}_i$ is determined (line 18), he will choose the network-times points based on his action $\boldsymbol{s}_i(\boldsymbol{\theta_i})$ (line 19). 
	
\end{itemize}

\subsubsection{Algorithm \ref{algo:ns_mo}}	\label{sec:algomo}
	
	Then, we describe how the operator can compute the aggregate network congestion statistics $p_{-i}[q,(n,t)]$ in Algorithm \ref{algo:ns_mo}.
	Based on the information $p_j[n,t]$ obtained from other users $j \in \mc{I} \backslash \{i\}$, the operator compute $p_{-i}[q,(n,t)]$ in \eqref{equ:pqnt} by counting the probabilities of the network selection with congestion level $q$ (excluding user $i$). We define $\mathbb{M}_{-i}(q)$ as the set of all possible subsets of set $\mc{I} \backslash \{i\}$ with $q$ users. As an example, for $\mc{I} = \{1,2,3,4\}$, we have $\mathbb{M}_{-3}(2) = \bigl\{ \{1,2\}, \{1,4\}, \{2,4\} \bigr\}$.

\begin{algorithm} [t] \small
\caption{\emph{Information update algorithm for the mobile operator.}}
\begin{algorithmic} [1] 

\STATEx \underline{Initialization}

\STATE Establish subscription plan database for users to retrieve: $\mathcal{M}[i,l,t]$ $\forall \, i \in \mathcal{I}, l \in \mathcal{L}, t \in \mathcal{T}$. 

\STATE Establish network parameter database for users to retrieve: Network capacity $\mu[n], \, \forall \, n \in \mathcal{N}$,  switching time $\delta[n,n']$, and switching cost $c[n,n'], \, \forall \, n,n' \in \mathcal{N}, n \neq n'$.

\STATE Allocate memory for the users' network selection statistics $\bs{p}_i, \, \forall \, i \in \mathcal{I}$.

\STATE Synchronize the clock timer $\tau := 1$ with all the users.

\STATEx \underline{Information Update for Bayesian NSG in Algorithm \ref{algo:ns_mu}} 

\STATE \textbf{repeat}

\STATE $ \ \ \ $ \textbf{if} network congestion update request is received from user $i$

\STATE $ \ \ \ \ \ $ Calculate the pmf $p_{-i}[q,(n,t)]$ of the congestion level \STATEx \quad\quad\quad $q = 0, \ldots, I-1$ for all user $i \in \mc{I}$ and $(n,t) \in \mathcal{N} \times \mathcal{T}$, \STATEx \quad\quad\quad and update in the database: 
\begin{equation} \label{equ:pqnt}
  p_{-i}[q,(n,t)] := \hspace{-0.3cm} \sum_{\mathcal{M} \in \mathbb{M}_{-i}(q)} \hspace{-0.1cm} \Bigl( \prod_{m \in \mathcal{M}} p_m[n,t] \Bigr) \! \Bigl( \hspace{-0.2cm} \prod_{k \in \mathcal{I} \backslash \mathcal{M} \cup \{i\}} \hspace{-0.3cm} \bigl(1 - p_k[n,t]\bigr) \Bigr). 
\end{equation}
%


\STATE $ \ \ \ $ \textbf{end if}

\STATE $ \ \ \ $ Set $\tau := \tau + 1$.

\STATE \textbf{until} $\tau \geq \tau^{\text{max}}$.

\end{algorithmic} \label{algo:ns_mo}
\end{algorithm}

\subsubsection{Computational Complexity}

  We are interested in understanding how long it takes to converge to a pure strategy BNE in the planning phase in Algorithm \ref{algo:ns_mu}. Theorem \ref{thm:complexity} ensures that each best response update (i.e., lines 7-13 in Algorithm \ref{algo:ns_mu}) can be computed efficiently in polynomial time.

\begin{theorem} \label{thm:complexity}
  Each best response update of user $i$ can be computed in $\mc{O}(|\Theta_i| N^3 T^3)$ time.
\end{theorem}

  The proof of Theorem \ref{thm:complexity} is given in Appendix \ref{app:complexity}. 
	For the number of best response updates required for convergence, we will study it in the next section  through numerical examples. 

\section{Performance Evaluations} \label{sec:pe}
 
  In this section, we study the performance of our DNS algorithm by comparing it with two benchmark schemes over various system parameters. 
	More specifically, to evaluate the users' level of satisfaction and to understand the network selections of these schemes, we evaluate the user utility, fairness, and the amount of network switching.
	We also show the impact of the prior probabilities of the mobility patterns on the network selections.

\subsection{Parameters and Settings}
  
  For each set of system parameters, we run the simulations $10000$ times with randomized network settings and users' mobility patterns in MATLAB and show the average value.
  Unless specified otherwise, the cellular (LTE category 5) network capacity $\mu[1]$ and the Wi-Fi network capacity $\mu[n], n \in \mathcal{N}_{\text{wifi}}$ are normally distributed random variables with means equal to $300$ Mbps \cite{wiki_eutra} and $54$ Mbps\footnote{Besides the difference in wireless communication standards, we consider a mean cellular network capacity much higher than the mean Wi-Fi network capacity, because the scale of the cellular network is larger and covers the users in multiple locations, while a Wi-Fi AP covers users in one particular location only.} \cite{ieee80211std}, respectively, with standard deviations equal to $5$ Mbps. 
  The probability that a Wi-Fi connection is available at a particular location $p^{\text{wifi}}$ is equal to $0.5$.
  We consider a two-minute duration, where the duration of a time slot $\Delta t = 10$ seconds, so $T = 12$.
	We assume that the network switching time is equal to one (i.e., $\delta[n,n'] = 1, \, \forall \, n,n' \in \mathcal{N}, n \neq n'$).
  For the switching cost not involving the idle network, we assume that they are the same that $c[n,n'] = c^{\textit{switch}}, \, \forall \, n,n' \in \mathcal{N} \backslash \{0\}, n \neq n'$.
	However, the switching cost involving the idle network is halved (i.e., $c[0,n] = c[n,0] = c^{\textit{switch}} / 2, \, \forall \, n \in \mathcal{N}$) such that \eqref{equ:triangleineq} in Assumption \ref{ass:idle}(c) is satisfied.	
  We consider that all the Wi-Fi networks are available to all the users within its coverage all the time.
  
	In our performance evaluation, we compare our DNS scheme against two benchmark schemes:
	
\begin{itemize}
	\item \emph{Cellular-only} scheme: The users use the cellular network all the time to avoid network switching, hence there is no data offloading.
	
	\item \emph{On-the-spot offloading} (OTSO) scheme: It is an offloading policy commonly used in most mobile devices today \cite{rayment_ac12}, where the data traffic is offloaded to Wi-Fi network whenever Wi-Fi is available. 
	Otherwise, if Wi-Fi networks are not available soon, the cellular connection will be used. 
	
\end{itemize}

\begin{figure*}[t]
\hspace{-0.5cm}
\centering
\begin{minipage}[t]{0.3\linewidth}
       \includegraphics[width=5.8cm, trim = 0.5cm 0cm 0cm 0cm, clip = true]{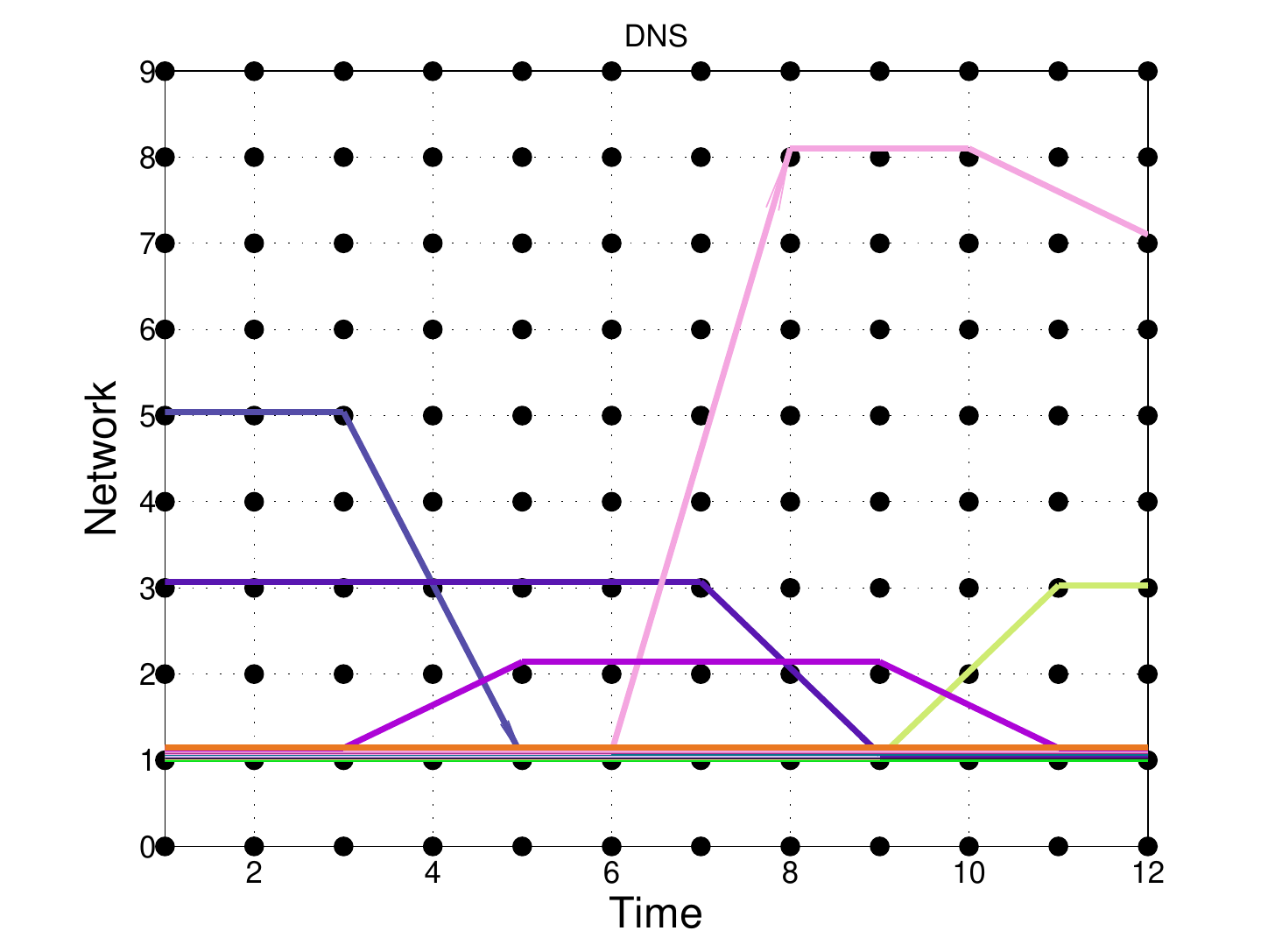}
   \label{fig:networksel_dns}
\end{minipage}
\quad
\begin{minipage}[t]{0.3\linewidth}
       \includegraphics[width=5.8cm, trim = 0.5cm 0cm 0cm 0cm, clip = true]{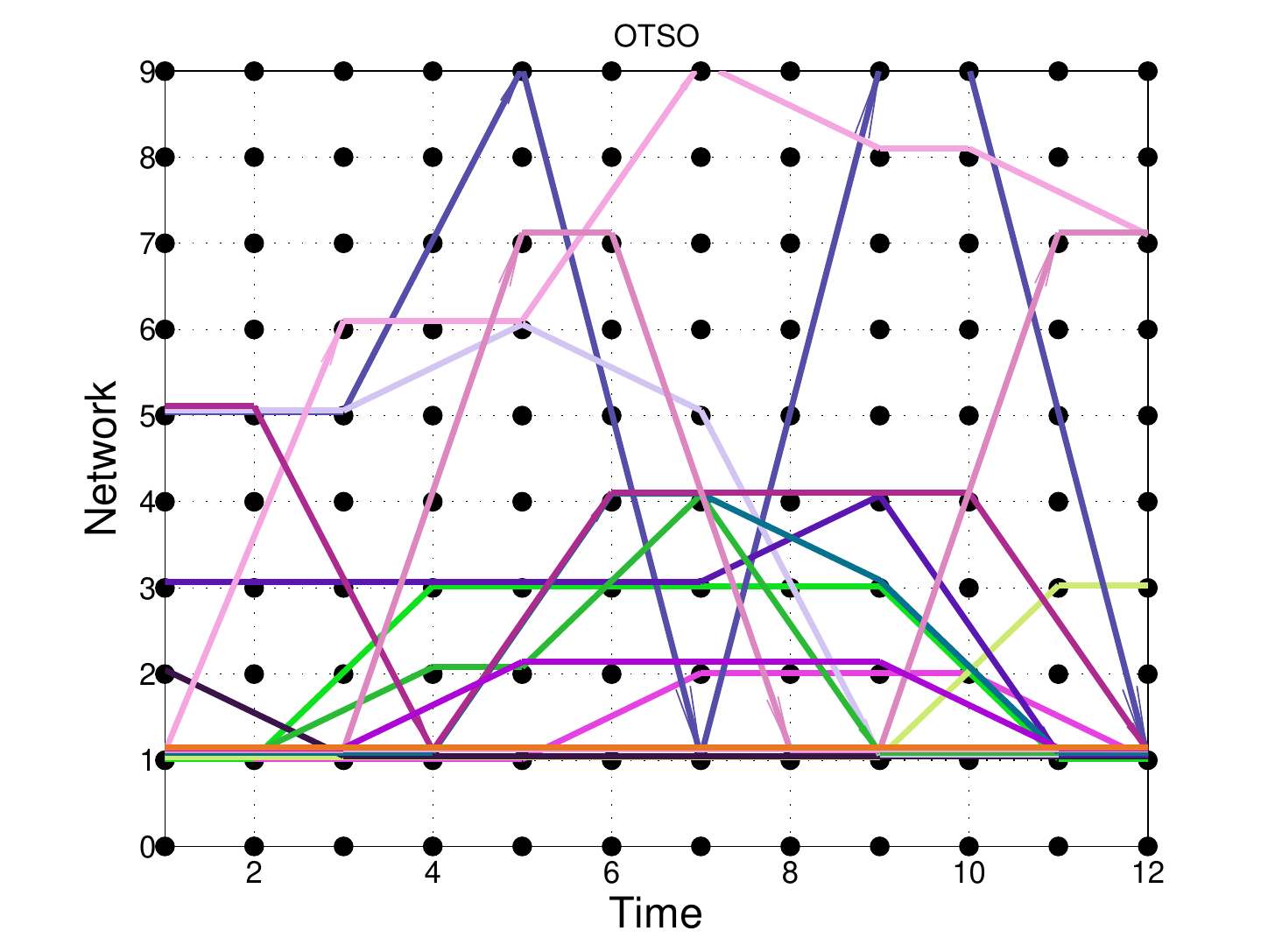}
   \label{fig:networksel_otso}
\end{minipage}
\quad
\begin{minipage}[t]{0.3\linewidth}
       \includegraphics[width=5.8cm, trim = 0.5cm 0cm 0cm 0cm, clip = true]{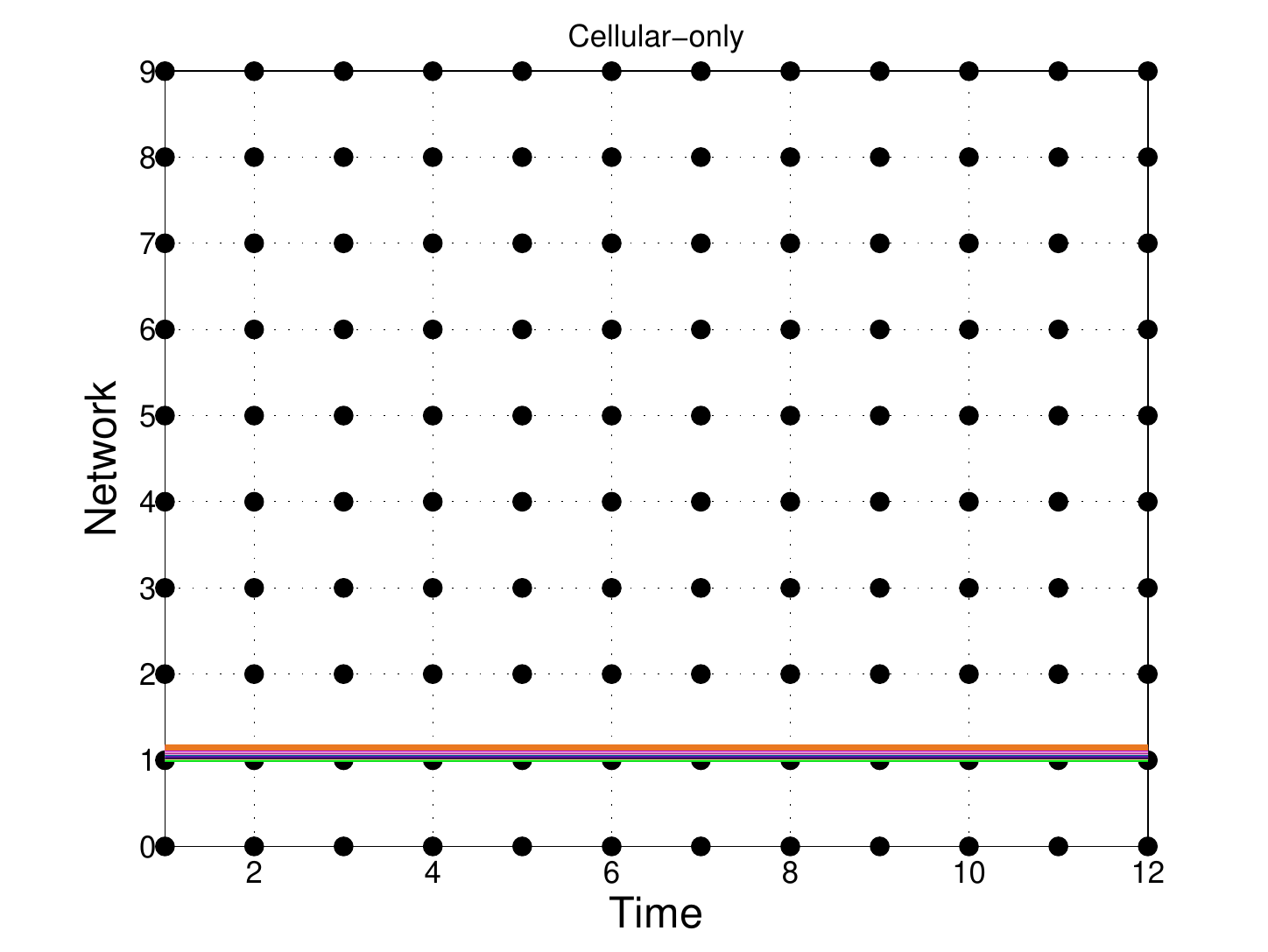} 
   \label{fig:networksel_cell}
\end{minipage}	
\vspace{-0.2cm}
\caption{Illustration of the network selection of DNS, OTSO, and cellular-only schemes with $I = 15$ users and $c^{\textit{switch}} = 400$.}
\label{fig:networksel}
\end{figure*} 
\begin{figure*}[t]
\hspace{-0.5cm}
\centering
\begin{minipage}[t]{0.3\linewidth}
       \includegraphics[width=5.8cm, trim = 0.5cm 0cm 0cm 0cm, clip = true]{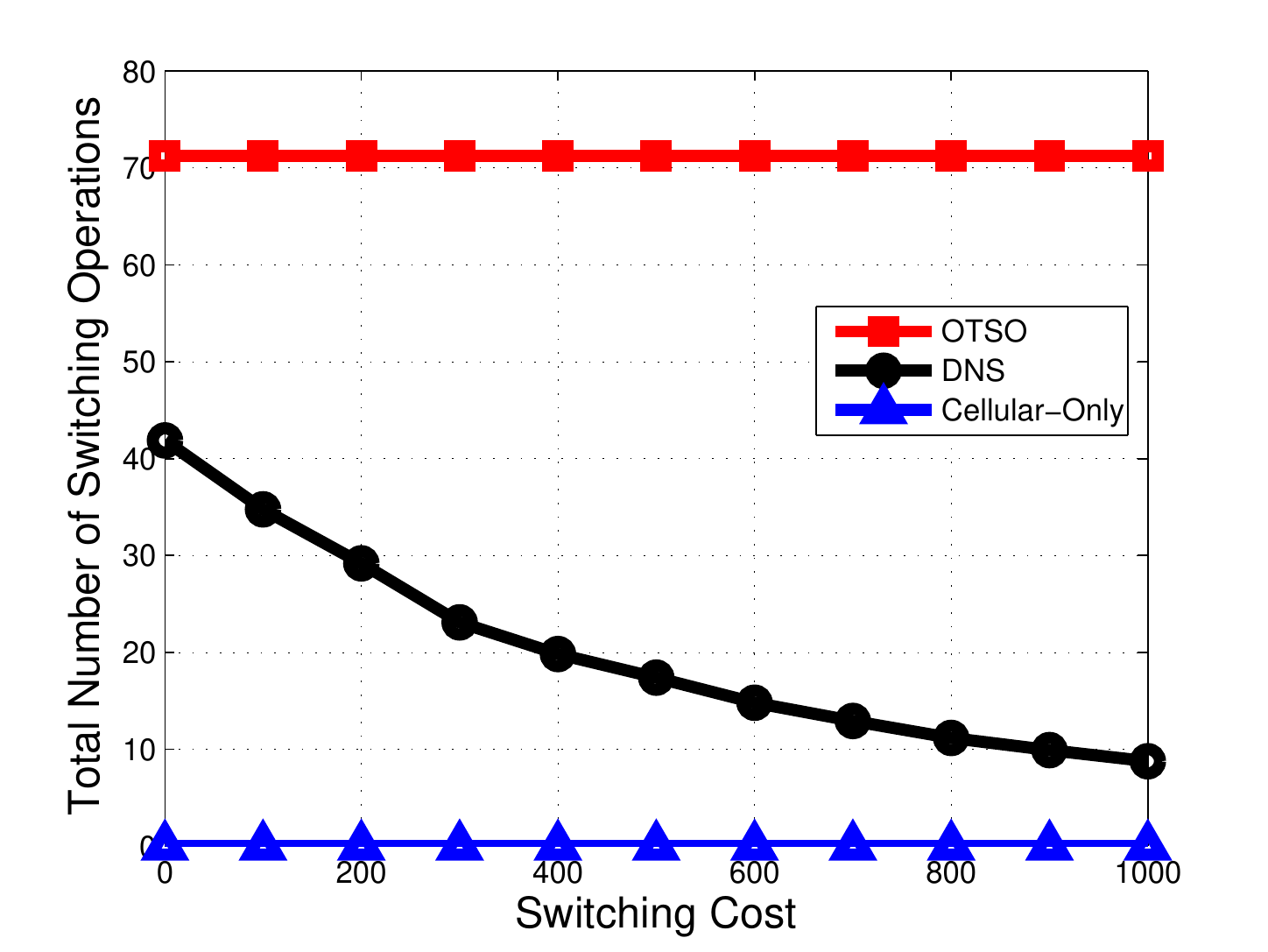} 
   \caption{The total number of switching operations versus switching cost $c^{\textit{switch}}$ for $I = 30$.}
   \label{fig:switches_swcost}
\end{minipage}
\quad
\begin{minipage}[t]{0.3\linewidth}
       \includegraphics[width=5.8cm, trim = 0.5cm 0cm 0cm 0cm, clip = true]{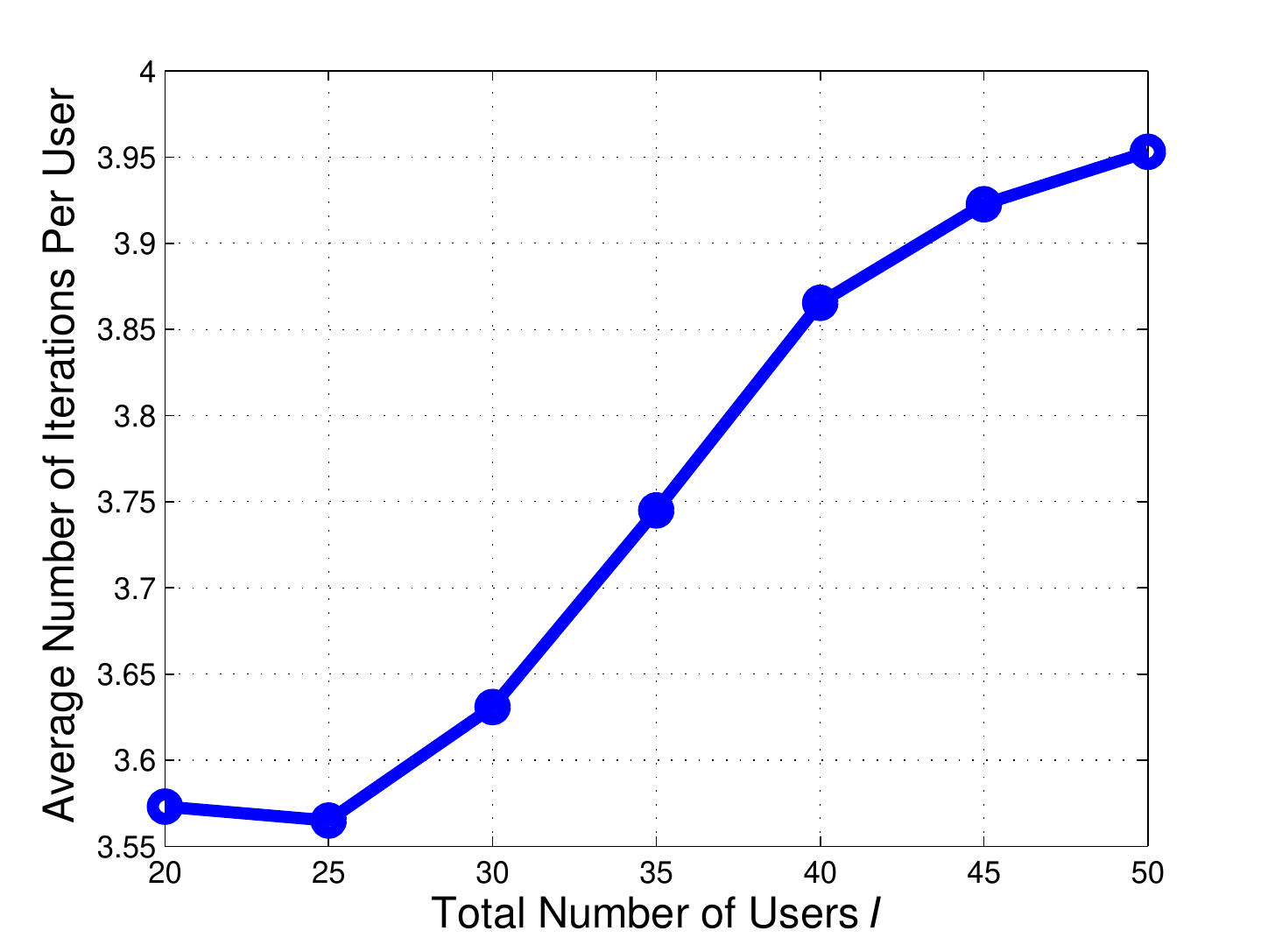} 
   \caption{The average number of best response update iterations per user for convergence in the DNS scheme with $c^{\textit{switch}} = 400$.}
   \label{fig:numbrperuser_numusers}
\end{minipage}	
\quad
\begin{minipage}[t]{0.3\linewidth}
       \includegraphics[width=5.8cm, trim = 0.4cm 0cm 0cm 0cm, clip = true]{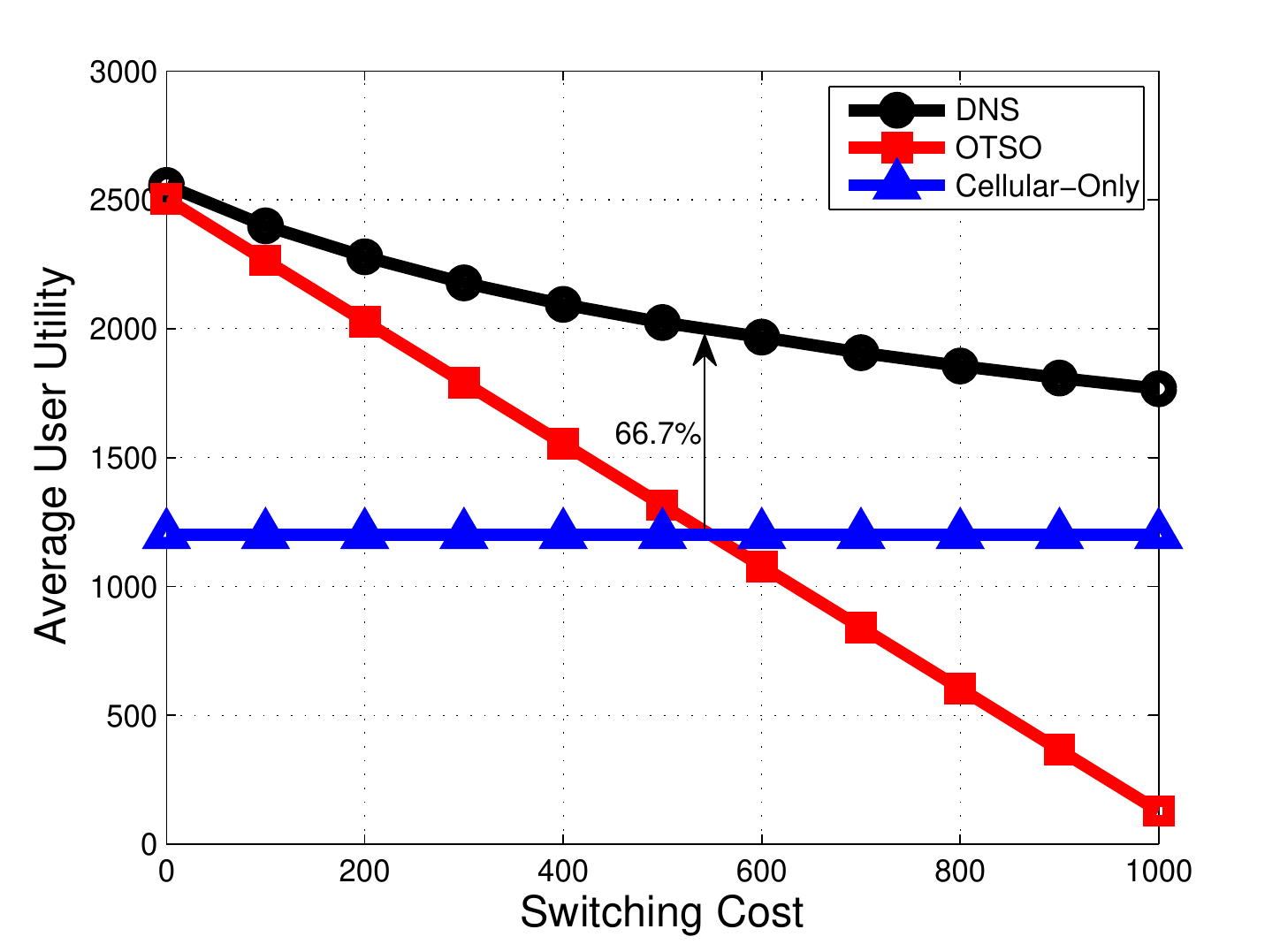} 
   \caption{The average user utility versus switching cost $c^{\textit{switch}}$ for $I = 30$.}
   \label{fig:utility_swcost}
\end{minipage}
\vspace{-0.2cm}
\end{figure*}


\subsection{Performance Evaluations of Deterministic Mobility Patterns} \label{sec:pe_deterministic}

	In this subsection, we first evaluate the schemes in the deterministic mobility pattern case.
	Here, the users' deterministic mobility patterns are generated based on the same location transition matrix $P = [p(l' \,|\, l)]_{L \times L}$, where $p(l' \,|\, l)$ is the probability that a user will move to location $l'$ given that it is currently at location $l$.\footnote{In our simulations, we just use the location transition matrix as a way to generate the users' mobility patterns. We want to clarify that it does not matter how to generate the users' locations as long as they are given as system parameters in this paper.} 
	All user move around $L = 16$ possible locations in a four by four grid (similar to that in Fig.~\ref{fig:network}).
	The probability that the user stays at a location is $p(l \,|\, l) = 0.6, \, \forall \, l \in \mathcal{L}$. Moreover, it is equally likely for the user to move to any one of the neighbouring locations. 
	Take location $7$ in Fig.~\ref{fig:network} as an example, the probability that the user will move to locations $3$, $6$, $8$, or $11$ is equal to $(1-0.6)/4 = 0.1$.  
  For the edge location $12$, however, the probability of  moving to any one of its three neighbouring locations is $(1-0.6)/3 = 0.133$.

\subsubsection{Illustration of Network Selection Schemes}

  In Fig.~\ref{fig:networksel}, we illustrate the network selections under DNS, OTSO, and cellular-only schemes for $I = 15$ users and a switching cost $c^{\textit{switch}} = 400$. 
	We can see that the OTSO scheme prefers Wi-Fi networks, so it results in a lot of network switching.
	On the other extreme, the cellular-only scheme uses only the cellular network, so there is no network switching.

\subsubsection{Network Switching and Scalability} 

  \textbf{(Summary of observations)} We first show that the DNS scheme is able to \emph{adaptively} choose the number of switching operations based on the switching cost. 
	We also show that the DNS scheme is \emph{scalable} by considering the number of best response updates for convergence.

	In Fig.~\ref{fig:switches_swcost}, we plot the total number of network switching operations against the switching cost $c^{\textit{switch}}$  for $I = 30$. 
	We can see that the performances of both the cellular-only and OTSO schemes are static as they are independent of $c^{\textit{switch}}$. However, the DNS scheme responds to the increasing switching cost by decreasing the number of switching.


  In Theorem \ref{thm:complexity}, we have established that each best response update can be computed in polynomial time.
  In Fig. \ref{fig:numbrperuser_numusers}, we continue with the evaluation of the convergence speed of Algorithm \ref{algo:ns_mu} by counting the average number of best response updates per user required for convergence with respect to different $I$ with $c^{\textit{switch}} = 400$.
  We observe that Algorithm \ref{algo:ns_mu} scales well with the increasing user population.
  In particular, each user only needs to perform $3.57$ and $3.95$ best response updates on average for $I = 20$ and $I = 50$, respectively, before the strategy profile converges to a pure strategy BNE.

\subsubsection{Average User Utility} 

  \textbf{(Summary of observations)} In this subsection, we study the impact of various system parameters on the user utility.  	  
	Overall, we find that the DNS scheme achieves the highest utility by taking into account both the \emph{ping-pong effect} and \emph{Wi-Fi network congestion}. 
	The results also reveal that the OTSO performs well under a low switching cost and a low Wi-Fi availability.

	In Fig.~\ref{fig:utility_swcost}, we plot the average user utility against the switching cost $c^{\textit{switch}}$ for $I = 30$.
	First, we observe that the proposed DNS scheme achieves the highest user utility compared with OTSO and cellular-only schemes. 
	More specifically, the DNS scheme improves the utility of these two schemes by $66.7\%$ when $c^{\textit{switch}} = 550$.
	In addition, for the DNS scheme, we see that its utility decreases gradually with $c^{\textit{switch}}$. This is because DNS is aware of the increasing switching cost and thus reduces the number of switching operations (as shown in Fig.~\ref{fig:switches_swcost}), which results in a milder reduction in utility.
	For the OTSO scheme, as it is unaware of the switching cost, the average user utility experiences a heavy reduction when $c^{\textit{switch}}$ is large.
	For the cellular-only scheme, since it does not perform any network switching, the user utility is independent of $c^{\textit{switch}}$.

\begin{figure*}[t]
\hspace{-0.5cm}
\centering
\begin{minipage}[t]{0.3\linewidth}
       \includegraphics[width=5.8cm, trim = 0.4cm 0cm 0cm 0cm, clip = true]{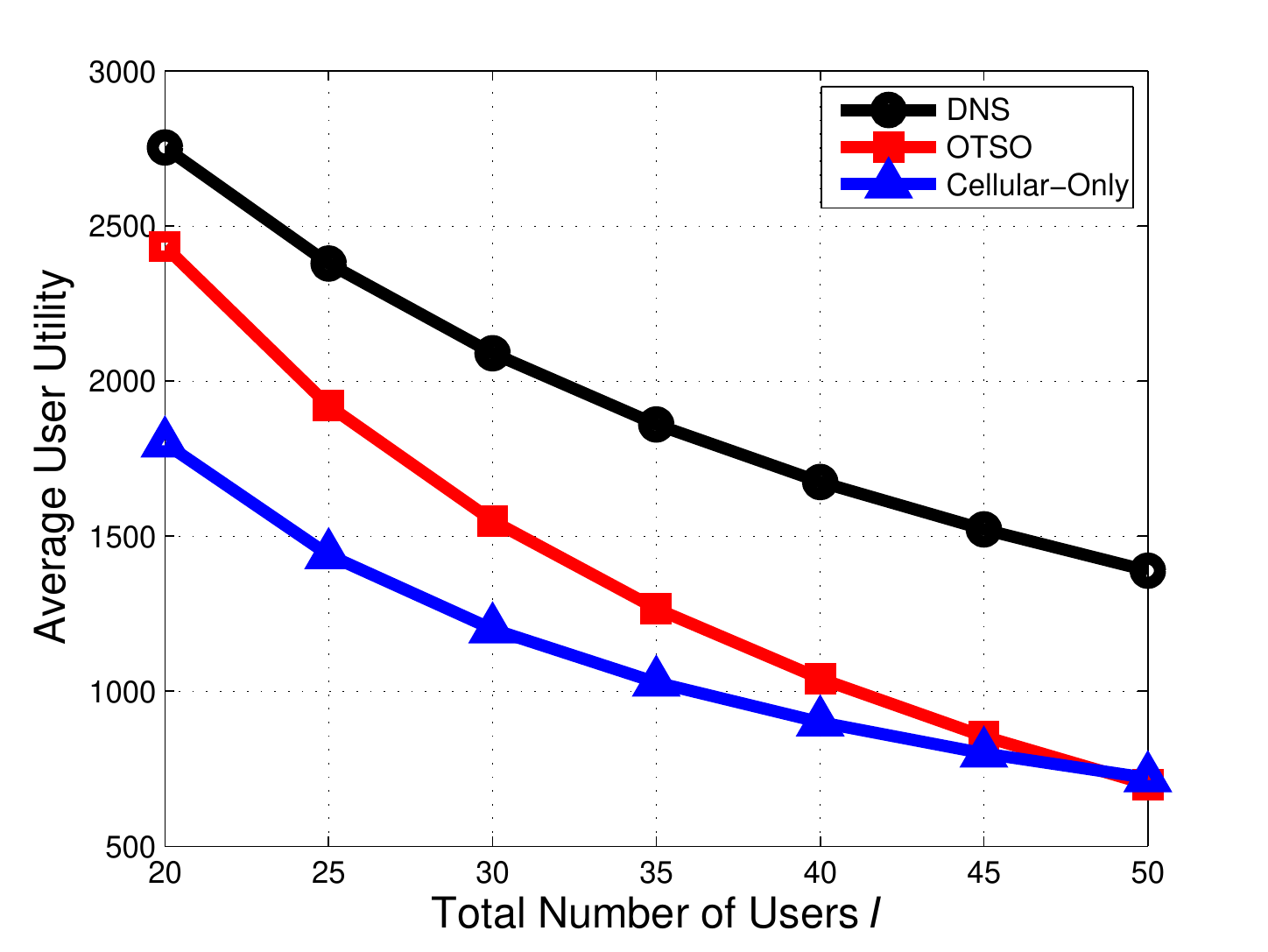} 
   \caption{The average user utility versus number of users $I$ for $c^{\textit{switch}} = 400$.}
   \label{fig:utility_numusers}
\end{minipage}
\quad
\begin{minipage}[t]{0.3\linewidth}
       \includegraphics[width=5.8cm, trim = 0.4cm 0cm 0cm 0cm, clip = true]{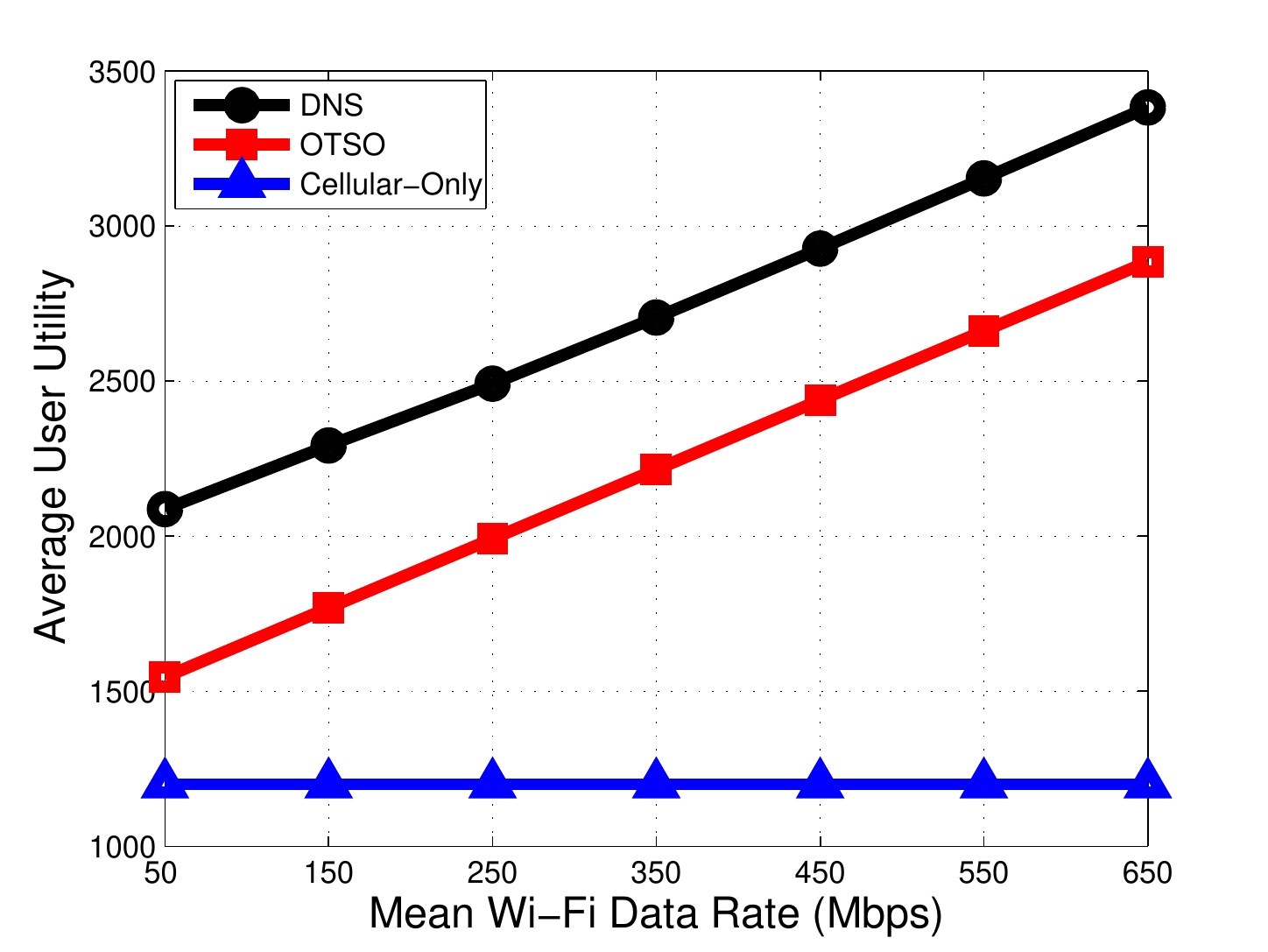} 
   \caption{The average user utility versus the mean Wi-Fi data rate for $I = 30$ and $c^{\textit{switch}} = 400$.}
   \label{fig:utility_wifirate}
\end{minipage}
\quad
\begin{minipage}[t]{0.3\linewidth}
       \includegraphics[width=5.8cm, trim = 0.4cm 0cm 0cm 0cm, clip = true]{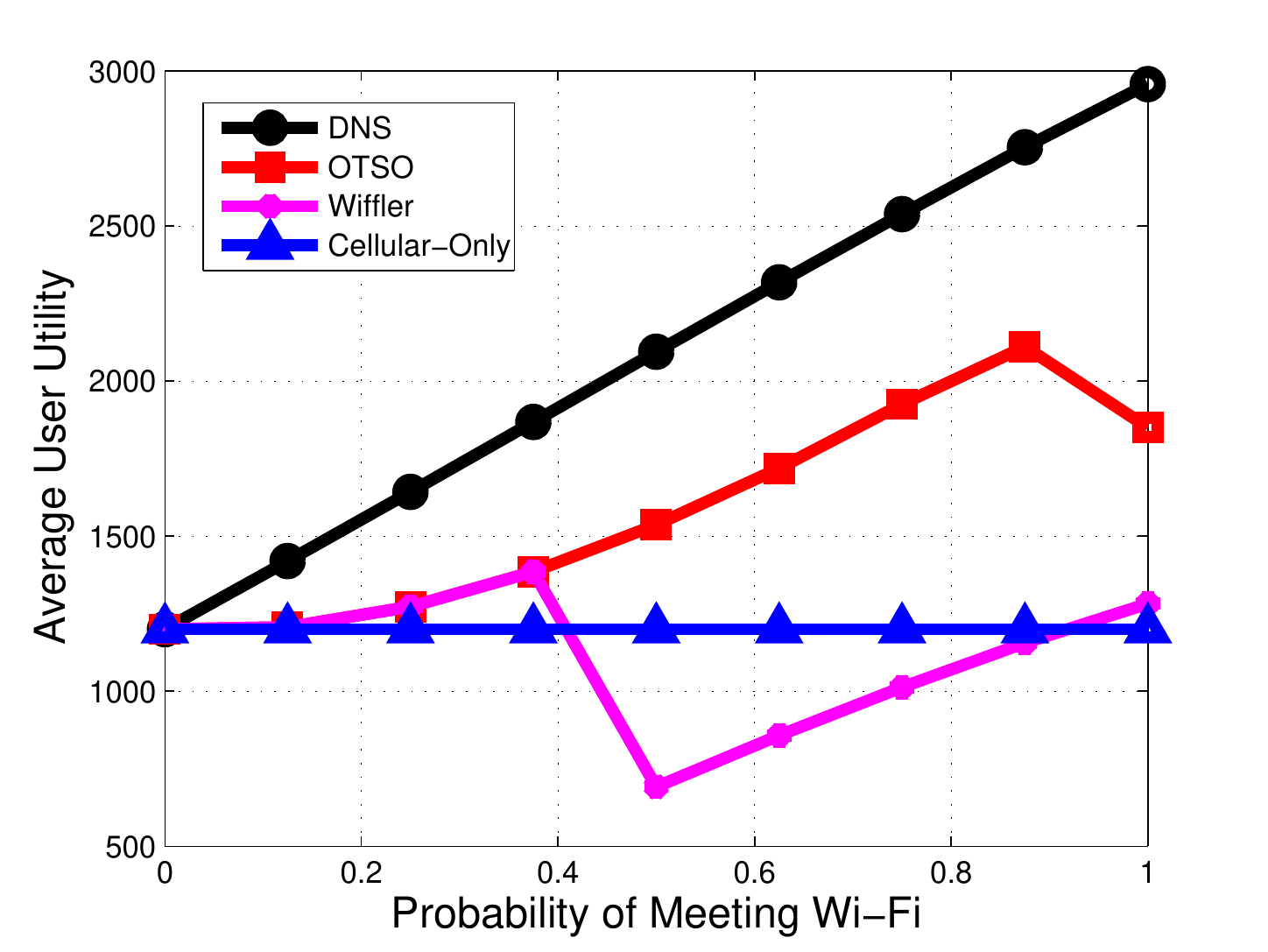}
   \caption{The average user utility versus the probability of meeting Wi-Fi for $I = 30$ and $c^{\textit{switch}} = 400$.}
   \label{fig:utility_probwifi}
\end{minipage}
\vspace{-0.2cm}
\end{figure*}

	In Fig.~\ref{fig:utility_numusers}, we plot the average user utility against the number of users $I$ for $c^{\textit{switch}} = 400$. 
  In general, when $I$ increases, the congestion level increases, so the average utility under all three schemes decrease.  
  We observe that the DNS scheme results in the highest user utility, which suggests that it achieves a good load balancing across the networks.
	For the cellular-only scheme, since it does not access any available Wi-Fi network capacity, the average user utility is significantly low.

	In Fig.~\ref{fig:utility_wifirate}, we plot the average user utility against the mean Wi-Fi data rate for $I = 30$ and $c^{\textit{switch}} = 400$. We observe that the result is intuitive, where the utility under both the DNS and OTSO schemes increases with the mean Wi-Fi data rate. Also, the DNS scheme achieves the highest user utility among the three schemes.
	
	Furthermore, we aim to study the impact of the probability of meeting Wi-Fi $p^{\text{wifi}}$ on different schemes. Here, we compare with an additional \emph{Wiffler} scheme \cite{balasubramanian_am10}, which is a prediction-based offloading scheme that operates as follows. Let $\zeta$ be the estimated amount of data that can be transferred using Wi-Fi by the deadline. 
  If Wi-Fi is available in the current location, then Wi-Fi will be used immediately. If Wi-Fi is not available, the user needs to check whether the condition $\zeta \geq \theta k$ is satisfied, where $k$ is the remaining size of the file to be transferred, and $\theta > 0$ is the conservative coefficient that tradeoffs the amount of data offloaded with the completion time of the file transfer. If this condition is satisfied, meaning that the estimated data transfer using Wi-Fi is large enough, then the user will stay idle and wait for the Wi-Fi connection. Otherwise, the user will use the cellular connection. Here, we set $\theta = 1$ as suggested in \cite{balasubramanian_am10} and consider $k = 0.5$ in the simulation.

	In Fig.~\ref{fig:utility_probwifi}, we plot the average user utility against the probability of meeting Wi-Fi $p^{\text{wifi}}$ for $I = 30$ and $c^{\textit{switch}} = 400$.
	For the DNS scheme, we can see that the utility increases with $p^{\text{wifi}}$, as the users experience a lower level of network congestion when more Wi-Fi networks are available.
	For the OTSO scheme, we observe a similar trend from small to medium $p^{\text{wifi}}$. Surprisingly, it experiences a drop in utility when $p^{\text{wifi}}$ is above $0.9$. The reason is that when $p^{\text{wifi}}$ is high such that Wi-Fi coverage is almost ubiquitous, all the users would use the Wi-Fi networks all the time, making the Wi-Fi networks very congested but leaving the cellular network with almost no user. Thus, the average user utility at $p^{\text{wifi}} = 1$ corresponds to the average throughput obtained from the Wi-Fi networks only (i.e., excluding the cellular network) minus the total switching cost.
	For the cellular-only scheme, since it is independent of the Wi-Fi availability, the average user utility is independent of $p^{\text{wifi}}$. 
	For the Wiffler scheme, when $p^{\text{wifi}} < 0.5$, it is the same as the OTSO scheme, which prefers to use Wi-Fi network when it is available, but the cellular network otherwise. However, when $p^{\text{wifi}} \geq 0.5$, it becomes a Wi-Fi only scheme, which it will remain idle (instead of using the cellular network) when Wi-Fi is not available. Thus, its user utility increases with $p^{\text{wifi}}$ when more Wi-Fi networks are available.

\begin{figure*}[t]
\hspace{-0.5cm}
\centering
\begin{minipage}[t]{0.3\linewidth}
       \includegraphics[width=5.8cm, trim = 0.5cm 0cm 0cm 0cm, clip = true]{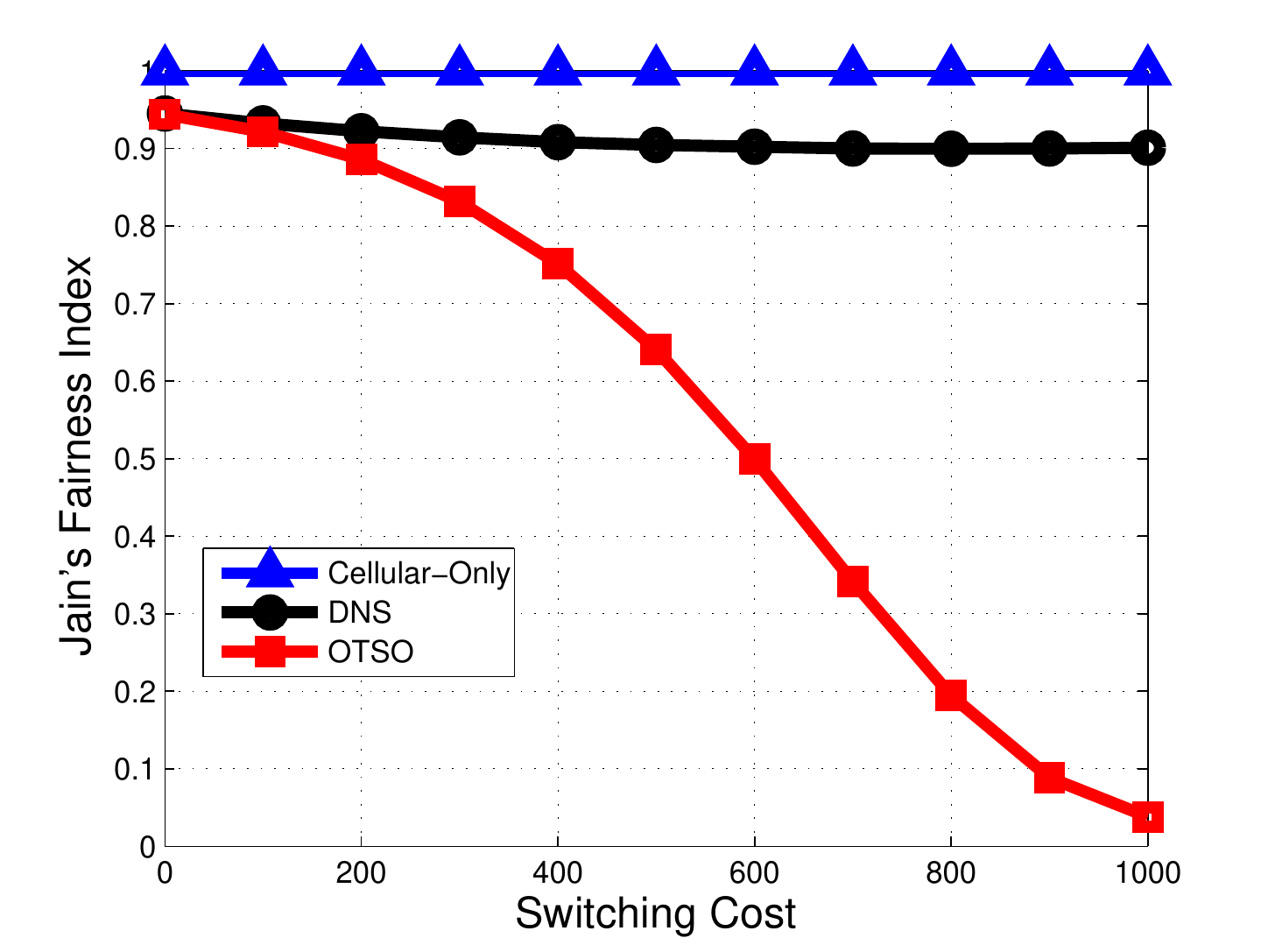} 
   \caption{The Jain's fairness index versus switching cost $c^{\textit{switch}}$ for $I = 30$.} 
   \label{fig:jains_swcost}
\end{minipage}
\quad
\begin{minipage}[t]{0.3\linewidth}
       \includegraphics[width=5.8cm, trim = 0.4cm 0cm 0cm 0cm, clip = true]{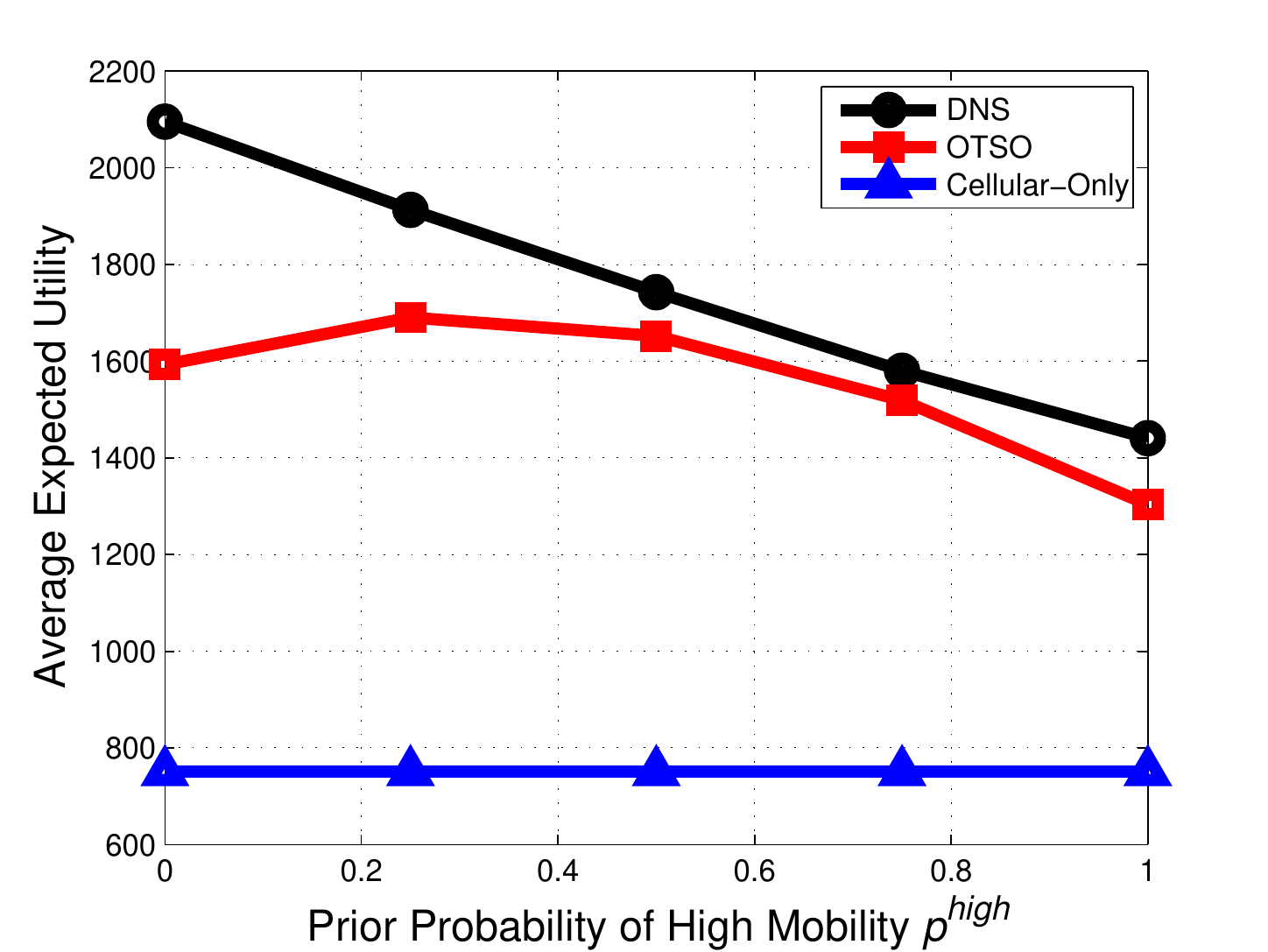} 
 \caption{The average user utility versus the prior probability $p^{\textit{high}}$ of high mobility for $I = 8$ and $c^{\textit{switch}} = 10$.}
\label{fig:utility_prior}
\end{minipage}
\quad
\begin{minipage}[t]{0.3\linewidth}
       \includegraphics[width=5.8cm, trim = 0.5cm 0cm 0cm 0cm, clip = true]{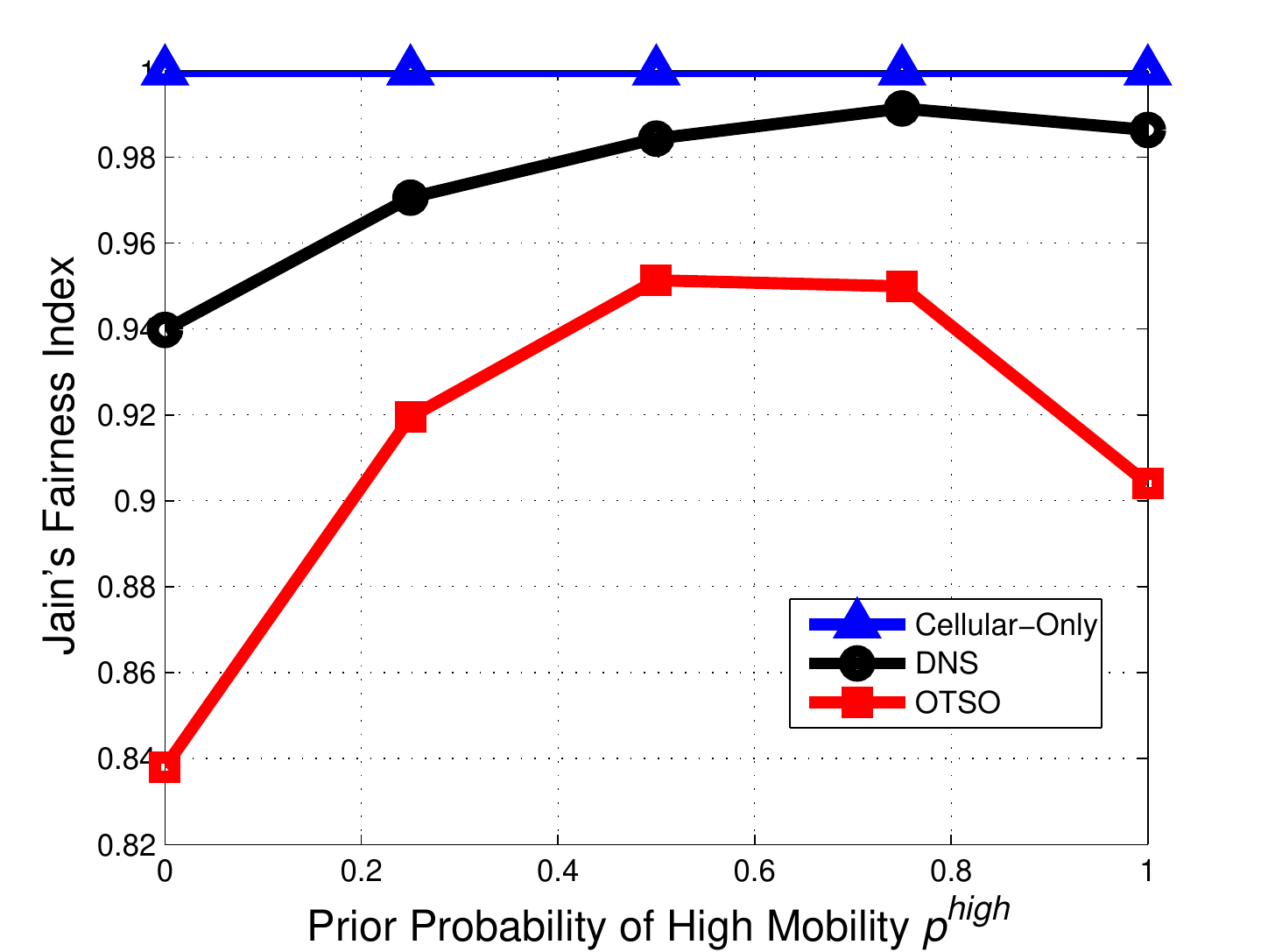} 
 \caption{The Jain's fairness index versus the prior probability $p^{\textit{high}}$ of high mobility for $I = 8$ and $c^{\textit{switch}} = 10$.}
\label{fig:jains_prior}
\end{minipage}
\vspace{-0.2cm}
\end{figure*}

\subsubsection{Fairness} 

  \textbf{(Summary of observations)} In this subsection, we study the fairness of the network resource allocation and show that the DNS scheme achieves a high degree of fairness. In addition, the fairness of the OTSO scheme decreases sharply under a high switching cost.

	In Fig.~\ref{fig:jains_swcost}, we evaluate the degree of fairness among the users by plotting the Jain's fairness index \cite{jain_aq84} defined as $\bigl( \sum_{i \in \mathcal{I}} U_i(\boldsymbol{r},\boldsymbol{\theta}) \bigr)^2 / \bigl( I \sum_{i \in \mathcal{I}} U_i(\boldsymbol{r},\boldsymbol{\theta})^2 \bigr)$ against $c^{\textit{switch}}$ for $I = 30$. 	
  Since the users under the cellular-only scheme always have the same utility, its fairness index is always equal to one.
  Furthermore, we notice that the fairness indices of both the DNS and OTSO schemes decrease with $c^{\textit{switch}}$. For the DNS scheme, as $c^{\textit{switch}}$ increases, the users switch networks less often (as shown in Fig.~\ref{fig:switches_swcost}). In this way, the utilities among the users at a larger $c^{\textit{switch}}$ are less balanced than that at a smaller $c^{\textit{switch}}$, so the degree of fairness decreases with $c^{\textit{switch}}$. 
  For the OTSO scheme, although its network selection is independent of $c^{\textit{switch}}$, the increase in $c^{\textit{switch}}$ widens the disparity in utilities among the users with different number of network switching.
  Moreover, we observe in Fig.~\ref{fig:jains_swcost} that the DNS scheme achieves a higher fairness index than the OTSO scheme, and the fairness index of the DNS scheme decreases much slowly than that of the OTSO scheme. It suggests that the adaptive DNS scheme results in a fairer resource allocation than the static OTSO scheme.

\subsection{Performance Evaluations of Random Mobility Patterns}

  In this subsection, we evaluate our proposed DNS scheme under the random mobility pattern case. 
	Here, we assume that the cellular network capacity $\mu[1]$ and the Wi-Fi network capacity $\mu[n], n \in \mathcal{N}_{\text{wifi}}$ are normally distributed random variables with means equal to $100$ Mbps and $50$ Mbps, respectively, and standard deviations equal to $5$ Mbps. 
	The probability of meeting Wi-Fi $p^{\text{wifi}} = 0.9$.
	The switching penalties are the same as that in Section \ref{sec:pe_deterministic}.
	We consider a one-minute duration, so $T = 6$ for $\Delta t = 10$ seconds.
	There are $I = 8$ users moving around $L = 5$ possible locations on a straight road.\footnote{Due to the relatively higher complexity to execute the DNS algorithm under the random mobility pattern case  (especially the need to run $1000$ times to have a good estimation of the average performance), we consider a smaller scale of simulation in this subsection.} 
  For each set of system parameters, we run the simulations $1000$ times with randomized network settings and users' mobility patterns in MATLAB and show the average value.


  For each user in the random mobility pattern case here, we consider that there are two possible mobility patterns that are generated with different characteristics: 

\begin{itemize}
	\item High mobility: With a prior probability $p^{\textit{high}}$, the user will \emph{frequently} move across $L$ locations. In the simulation, we assume that the user has a total probability of $0.9$ in moving to one of his neighboring locations and a probability of $0.1$ in staying at his current location.
	
	\item Low mobility: With a prior probability $1 - p^{\textit{high}}$, the user moves much less frequently. In the simulation, we assume that the user has a total probability of $0.1$ in moving to one of his neighboring locations and a probability of $0.9$ in staying at his current location.
\end{itemize}

\subsubsection{Impact of Prior Distribution of Mobility Patterns} \label{sec:prior}

  \textbf{(Summary of observations)} Consistent with the observations under the deterministic mobility case, we see that the DNS achieves the highest expected utility and a high level of fairness under the random mobility case.

	In Fig.~\ref{fig:utility_prior}, we plot the average expected utility against the prior probability $p^{\textit{high}}$ of high mobility when switching cost $c^{\textit{switch}} = 10$. 
	First, we can see that both the utilities under the DNS and OTSO schemes decrease with $p^{\textit{high}}$,  because of the higher total switching cost when the users switch networks more often under a high mobility.
	Nevertheless, the DNS scheme results in a higher expected user utility.
	For the cellular-only scheme, since the users select the cellular network regardless of their mobility, the user utility is independent of $p^{\textit{high}}$.

	In Fig.~\ref{fig:jains_prior}, we plot the Jain's fairness index \cite{jain_aq84} $\bigl( \sum_{i \in \mathcal{I}} EU_i(\boldsymbol{s}) \bigr)^2 / \bigl( I \sum_{i \in \mathcal{I}} EU_i(\boldsymbol{s})^2 \bigr)$ against $p^{\textit{high}}$ for $c^{\textit{switch}} = 10$. 
	We can see that the DNS scheme achieves a higher degree of fairness than the OTSO scheme.
	Moreover, for both the DNS and OTSO schemes, there is an increase in fairness when $p^{\textit{high}}$ increases from a small value to a medium value. We observe that it is due to the larger percentage drop in the expected utility for high-utility users, which increases the fairness.
	However, for the OTSO scheme, there is a further drop in fairness when $p^{\textit{high}} > 0.5$, because the expected utilities of some users (not necessarily the high-utility users) decrease, which leads to a reduction in fairness.
\section{Conclusions and Future Work} \label{sec:concl}

  In this paper, we studied the intelligent network selection problem with the objective of achieving an effective data offloading for cellular and Wi-Fi integration. In particular, we focused on understanding the impact of \emph{network congestion} and \emph{switching penalty} due to the herd behaviour and ping-pong effect, respectively, which were not systematically considered in the literature.
	As a benchmark, we formulated the centralized user utility maximization problem and showed that it is an NP-hard problem, which motivated us to consider a distributed approach.
	More specifically, with the statistical information of the user mobility, we formulated the users' interactions as a Bayesian network selection game, proved that it is a potential game, and proposed a distributed network selection (DNS) algorithm with provably nice convergence properties.
	Compared with the on-the-spot offloading (OTSO) and cellular-only schemes, our simulation results showed that the proposed DNS algorithm results in the highest user utility and a good fairness by avoiding Wi-Fi network congestion and costly network switching.	
	 In addition, we showed that the OTSO scheme performs especially well under a low switching cost and a low Wi-Fi availability.
	
  In this work, we considered the static setting where each user knows the network conditions and the statistical information of his possible mobility patterns. 
	For the future work, we plan to consider a dynamic setting where a user needs to make online network selections, while considering the time-varying network conditions and mobility patterns. 
	Moreover, we have remarked that the complexity of implementing the DNS algorithm in the random mobility pattern case can be high. Thus, it is important to design a low-complexity DNS algorithm to converge to an approximate equilibrium of the game, while still taking into account the network congestion, switching penalty, and user mobility that we considered in this paper. 
	In addition, it is interesting to analyze the performance of the proposed scheme under the framework of stochastic geometry \cite{bao_sg15}.

\appendix

\subsection{Proof of Lemma \ref{lem:socialwelware}} \label{app:socialwelware}

  We prove the lemma by contradiction. 
	Assume on the contrary that for a socially optimal action profile $\bs{r}^*$, there exists $(n,t) \in \mc{N} \times \mc{T}$ such that $\omega[(n, t), \boldsymbol{r}^*, \boldsymbol{\theta}] > 1$.
	Without loss of generality, in $\bs{r}^*$, we assume that there exists user $i \in \mc{I}$ with route	
\begin{equation}
\begin{split}
	\boldsymbol{r}_i^* = \Bigl((n_i^{1},t_i^{1}),\ldots,(n_i^{q-1},t_i^{q-1}),(n_i^{q},t_i^{q}),(n_i^{q+1},t_i^{q+1}),\ldots,\\(n_i^{Q_i},t_i^{Q_i})\Bigr),
\end{split}
\end{equation}
where $\omega[(n_i^q, t_i^q), \boldsymbol{r}^*, \boldsymbol{\theta}] > 1$.
  We want to show that we can always find another action profile $\bs{r}'$ such that $\mathbb{W}(\boldsymbol{r}^*, \boldsymbol{\theta}) \leq \mathbb{W}(\boldsymbol{r}', \boldsymbol{\theta})$. 
	
	To do this, we define another action profile $\bs{r}'$, where user $i$ chooses \emph{not} to take the network-time point $(n_i^q, t_i^q)$ but remains idle. The new route taken by user $i$ is 
\begin{equation}
\begin{split}
	\boldsymbol{r}_i' = \Bigl((n_i^{1},t_i^{1}),\ldots,(n_i^{q-1},t_i^{q-1}), \hspace{4cm} \\
	\underbrace{(0, t_i^{q-1} + \delta[n_i^{q-1},0] + 1), \ldots,(0,t_i^{q+1} - \delta[0,n_i^{q+1}] - 1)}_{\text{remains idle}}, \\ (n_i^{q+1},t_i^{q+1}), \ldots,(n_i^{Q_i},t_i^{Q_i})\Bigr),
\end{split}
\end{equation}
which is a feasible network-time route (defined in Definition \ref{def:networktimeroute}) by Assumption \ref{ass:idle}(b).
However, all the other users choose the same route as they did in $\bs{r}^*$, i.e., $\bs{r}_j' = \bs{r}_j^*$ for all $j \in \mathcal{I} \backslash \{i\}$.

  Given the users' deterministic mobility patterns $\boldsymbol{\theta}$, we define the social welfare under action profile $\boldsymbol{r}$ as
\begin{equation} \label{equ:surplus}
\begin{split}
	\mathbb{W}(\boldsymbol{r},\boldsymbol{\theta})
	= \text{benefit}(\boldsymbol{r},\boldsymbol{\theta}) - \text{cost}(\boldsymbol{r}) \hspace{0.9cm} \\
	= \text{benefit}(\boldsymbol{r},\boldsymbol{\theta}) - \sum_{i \in \mathcal{I}} \text{cost}_i(\bs{r}_i),
\end{split}	
\end{equation}
where
\begin{equation}
	\mathbb{W}(\boldsymbol{r},\boldsymbol{\theta}) 
	\triangleq \sum_{i \in \mc{I}} U_i(\boldsymbol{r},\boldsymbol{\theta}), 
\end{equation}
\begin{equation} \label{equ:benefit2}
	\text{benefit}(\boldsymbol{r},\boldsymbol{\theta}) \triangleq  \sum_{i \in \mc{I}} \sum_{(n,t) \in \mc{V}(\boldsymbol{r}_i)} \frac{\mu[n]}{\omega[(n, t), \boldsymbol{r}, \boldsymbol{\theta}]}, 
\end{equation}
and
\begin{equation}
	\text{cost}_i(\bs{r}_i) = \sum_{\boldsymbol{e} \in \mathcal{E}(\boldsymbol{r}_i)} g[\boldsymbol{e}]. 
\end{equation}

	First, notice that we can express the total benefit in \eqref{equ:benefit2} as
\begin{equation} \label{equ:benefit}
	\text{benefit}(\boldsymbol{r}, \boldsymbol{\theta}) = \sum_{t \in \mathcal{T}} \sum_{n \in \mathcal{N}} \textbf{1}_{\{\omega[(n, t), \boldsymbol{r}, \boldsymbol{\theta}] \geq 1 \}} \mu[n],
\end{equation}
where $\textbf{1}_{\{.\}}$ is the indicator function.
  Since the set of network-time points with at least one user under $\boldsymbol{r}$ is the same as that under $\boldsymbol{r}^*$,  we have from Assumption \ref{ass:idle}(a) and \eqref{equ:benefit} that 
\begin{equation} \label{equ:benefit_derive}
	\text{benefit}(\boldsymbol{r}^*, \boldsymbol{\theta}) = \text{benefit}(\boldsymbol{r}', \boldsymbol{\theta}).
\end{equation}

	Second, from Assumption \ref{ass:idle}(c), we have for user $i$ that
\begin{equation} \label{equ:costj_derive}
	\text{cost}_i(\bs{r}_i^*) < \text{cost}_i(\bs{r}_i').
\end{equation}

	Third, the switching costs of other users remain the same such that
\begin{equation} \label{equ:costi_derive}
	\text{cost}_j(\bs{r}_j^*) = \text{cost}_j(\bs{r}_j'), \, \forall \, j \in \mathcal{I} \backslash \{i\}.
\end{equation}

  Overall, substituting \eqref{equ:benefit_derive}, \eqref{equ:costj_derive}, and \eqref{equ:costi_derive} into \eqref{equ:surplus}, we have 
\begin{equation} 
	\mathbb{W}(\boldsymbol{r}^*, \boldsymbol{\theta}) < \mathbb{W}(\boldsymbol{r}', \boldsymbol{\theta}),
\end{equation}
which leads to a contradiction.	\hfill \IEEEQED 

\subsection{Proof of Theorem \ref{thm:np_hard}} \label{app:np_hard}

  We prove the NP-hardness by \emph{restriction} \cite{garey_ca79}: We show that finding the social welfare maximization solution in a special case of a NSG can be transformed into a 3-dimensional matching decision problem, which is NP-complete \cite{garey_ca79, kleinberg_ad05}. 
	
	First, we define the 3-dimensional matching and its corresponding decision problem.

\begin{definition}[3-dimensional matching] \label{def:3dmatch} 
  Let $\mathcal{X}$, $\mathcal{Y}$, and $\mathcal{Z}$ be three finite disjoint sets. Let $\mathcal{R} \subseteq \mathcal{X} \times \mathcal{Y} \times \mathcal{Z}$ be a set of ordered triples, i.e., $\mathcal{R} = \{ (x,y,z): x \in \mathcal{X}, y \in \mathcal{Y}, z \in \mathcal{Z} \}$. Hence $\mathcal{R}' \subseteq \mathcal{R}$ is a \emph{3-dimensional matching} if for any two different triples $(x_1, y_1, z_1) \in \mathcal{R}'$ and $(x_2, y_2, z_2) \in \mathcal{R}'$, we have $x_1 \neq x_2$, $y_1 \neq y_2$, and $z_1 \neq z_2$.
\end{definition}

\begin{definition}[3-dimensional matching decision problem] \label{def:3dmatchproblem}
	Suppose $|\mathcal{X}| = |\mathcal{Y}| = |\mathcal{Z}| = N$. Given an input $\mathcal{R}$ with $|\mathcal{R}| \geq N$, decide whether there exists a 3-dimensional matching $\mathcal{R}' \subseteq \mathcal{R}$ with the maximum size $|\mathcal{R}'| = N$.
\end{definition}

  Consider a restricted NSG with the following restrictions, which is a special case of a NSG:	
	
(a) Networks and time slots: We consider that there are $T = 3$ time slots and $N$ available networks, and we do not consider the idle network. We assume that these $N$ networks are available to all the users at every location and time slot. Sets $\mathcal{X}$, $\mathcal{Y}$, and $\mathcal{Z}$ represent the sets of available networks in the three time slots, respectively, where $|\mathcal{X}| = |\mathcal{Y}| = |\mathcal{Z}| = N$.

(b) Network-time route: Set $\mathcal{R}$ represents the set of feasible network-time routes of all the users, i.e., $\mathcal{R}_i(\bs{\theta}_i) = \mathcal{R}, \, \forall \, i \in \mathcal{I}$. 
	Assume that each user $i$ can only choose one particular feasible network-time route $\bs{r}_i \in \mathcal{R}$.
  We assume that the number of users $I$ is large enough, such that the network-time routes of all the users cover all the network-time points in $\mathcal{X} \cup \mathcal{Y} \cup \mathcal{Z}$, so $|\mathcal{R}| = I \geq N$.
	Consider $\mathcal{R}' \subseteq \mathcal{R}$ in Definition \ref{def:3dmatchproblem} that represents a feasible network allocation. 
	We assume that a user, whose route is not chosen in $\mathcal{R}'$, will remain idle all the time and does not access any network.
		
(c) High network capacity: The benefit of using a network without any contention is larger than the switching cost to the network, i.e., $\mu[n] > c[n',n]$ for all networks $n,n' \in \mathcal{N}$. 

(d) Switching cost and switching time: The switching cost does not depend on the user's initial network configuration, i.e., $c[n,n'] = \tilde{c}[n']$ for all $n,n' \in \mathcal{N}$, where $\tilde{c}[n']$ is the switching cost to network $n' \in \mathcal{N}$. 
  In other words, the switching costs from network $n$ to a particular network $n'$ for all $n \in \mc{N}$ are the same.
	Also, the switching time between any two networks is zero. That is, $\delta[n,n'] = 0, \fa{n,n'}{N}$.

\begin{figure}[t]
 \centering
   \includegraphics[width=9cm, trim = 4cm 3.5cm 4cm 4cm, clip = true]{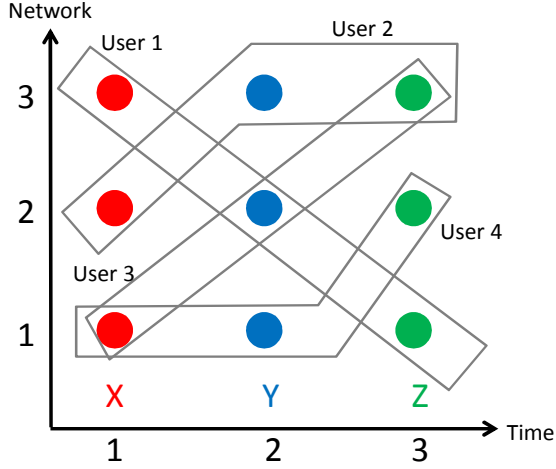} 
 \caption{An example of 3-dimensional matching. Here, the set $\mathcal{R}$ consists of the four grey areas, which represent the routes of users $1$ to $4$. The solution of the 3-dimensional matching problem is the set $\mathcal{R}'$ that consists of the routes of users $1$, $2$, and $4$. It is also the solution of the social welfare maximization problem.}
\label{fig:3dmatch}
\end{figure}

	In the restricted NSG, restrictions (c) and (d) imply that we can maximize the aggregate utility by covering \emph{all} the network-time points with \emph{any} available users. 
  Furthermore, Lemma \ref{lem:socialwelware} implies that we can focus on an optimal solution, where each network-time point should be chosen by \emph{at most one} user.
 So the optimal network allocation should not contain any overlapping components (i.e., multiple users choosing the same network-time point) as defined in Definition \ref{def:3dmatch}. 
	Putting the above discussions together, we know that in the aggregate utility maximization solution, \emph{every} element of $\mathcal{X} \times \mathcal{Y} \times \mathcal{Z}$ (i.e., every network) should be contained in \emph{exactly one} of the triples (i.e., the network-time routes) in $\mathcal{R}'$. 
	In other words, $\mathcal{R}' \subseteq \mathcal{R}$ is the optimal network allocation. So the social welfare maximization problem can be transformed to a 3-dimensional matching decision problem, which is NP-complete \cite{garey_ca79, kleinberg_ad05}. 
	By restriction, we establish that the problem of finding the social welfare maximization solution of the NSG is NP-hard.  
	An example is given in Fig.~\ref{fig:3dmatch}. \hfill \IEEEQED 

\subsection{Proof of Theorem \ref{thm:bpg}} \label{app:bpg}

  In the proof, we want to show that the utility function in \eqref{equ:utility_route} and the potential function in \eqref{equ:potential_deterministic} satisfy \eqref{equ:potential_deterministic_def}.
  First, starting from the original action profile $\bs{r} = (\bs{r}_i, \bs{r}_{-i})$, we define a new action profile $\bs{r}'$, where $\bs{r}_j' = \bs{r}_j$ if $j \neq i$ and $\bs{r}_j' \neq \bs{r}_j$ if $j = i$. In other words, only user $i$ changes its action from $\bs{r}_i$ to $\bs{r}_i'$ in the new action profile $\bs{r}' = (\bs{r}_i', \bs{r}_{-i})$. 
	
	Next, we define an partition of set $\mathcal{N} \times \mathcal{T}$, which consists of four non-overlapping sets of the network-time points
\begin{equation} 
\begin{split}
	\mc{B}^{(1)} = \{ (n,t): (n,t) \in \mc{V}(\bs{r}_i), (n,t) \notin \mc{V}(\bs{r}_i') \}, \\
	\mc{B}^{(2)} = \{ (n,t): (n,t) \notin \mc{V}(\bs{r}_i), (n,t) \in \mc{V}(\bs{r}_i') \}, \\
	\mc{B}^{(3)} = \{ (n,t): (n,t) \in \mc{V}(\bs{r}_i), (n,t) \in \mc{V}(\bs{r}_i') \}, \\
	\mc{B}^{(4)} = \{ (n,t): (n,t) \notin \mc{V}(\bs{r}_i), (n,t) \notin \mc{V}(\bs{r}_i') \},
\end{split}
\end{equation}
where $\mathcal{B}^{(1)} \cup \mathcal{B}^{(2)} \cup \mathcal{B}^{(3)} \cup \mathcal{B}^{(4)} = \mc{N} \times \mc{T}$.	
  As a result, considering the difference in congestion level in network-time point $(n,t)$ between action profiles $\boldsymbol{r}$ and $\boldsymbol{r}'$, we have
\begin{equation} \label{equ:change_congestion}
	\omega[(n,t), \boldsymbol{r}, \boldsymbol{\theta}] - \omega[(n,t), \boldsymbol{r}', \boldsymbol{\theta}] = 
\begin{cases} 
	  1, & \hspace{-0.3cm} \mbox{if } (n,t) \in \mc{B}^{(1)}, \\ 
	  -1, & \hspace{-0.3cm} \mbox{if } (n,t) \in \mc{B}^{(2)}, \\ 
	  0, & \hspace{-0.3cm} \mbox{if } (n,t) \in \mc{B}^{(3)} \cup \mc{B}^{(4)}. \\ 	
\end{cases} 
\end{equation}
	For example, in the first line in \eqref{equ:change_congestion}, we have one more user (i.e., user $i$) choosing the network-time point $(n,t) \in \mc{B}^{(1)}$ in the action profile $\bs{r} = (\bs{r}_i, \bs{r}_{-i})$ than in $\bs{r}' = (\bs{r}_i', \bs{r}_{-i})$, since users other than $i$ choose the same action profile $\bs{r}_{-i}$.
	
	Let $A \triangleq \sum_{\boldsymbol{e} \in \mathcal{E}(\boldsymbol{r}_i')} g[\boldsymbol{e}] - \sum_{\boldsymbol{e} \in \mathcal{E}(\boldsymbol{r}_i)} g[\boldsymbol{e}]$.
	As a result, we have
\begin{align} 
   & \Phi(\boldsymbol{r}_i, \boldsymbol{r}_{-i}, \boldsymbol{\theta}) - \Phi(\boldsymbol{r}_i', \boldsymbol{r}_{-i}, \boldsymbol{\theta}) \nonumber  \\ 
= & \Bigl(\sum_{j \in \mathcal{I}} \sum_{\boldsymbol{e} \in \mathcal{E}(\boldsymbol{r}_j')} g[\boldsymbol{e}] - \sum_{j \in \mathcal{I}} \sum_{\boldsymbol{e} \in \mathcal{E}(\boldsymbol{r}_j)} g[\boldsymbol{e}] \Bigr) \nonumber \\ 
	& +  \sum_{(n,t) \in \mathcal{N} \times \mathcal{T}} \Bigl( \sum_{q=1}^{\omega[(n,t), \boldsymbol{r}, \boldsymbol{\theta}]} \frac{\mu[n]}{q} - \sum_{q=1}^{\omega[(n,t), \boldsymbol{r}', \boldsymbol{\theta}]} \frac{\mu[n]}{q}  \Bigr) \nonumber \\ 
=  & A + \hspace{-0.3cm} \sum_{(n,t) \in \mathcal{B}^{(1)} \cup \mathcal{B}^{(2)} \cup \mathcal{B}^{(3)} \cup \mathcal{B}^{(4)}} \hspace{-0.3cm} \Bigl( \sum_{q=1}^{\omega[(n,t), \boldsymbol{r}, \boldsymbol{\theta}]} \frac{\mu[n]}{q} - \hspace{-0.3cm} \sum_{q=1}^{\omega[(n,t), \boldsymbol{r}', \boldsymbol{\theta}]} \frac{\mu[n]}{q} \Bigr) \nonumber \\ 
=  & A + \sum_{(n,t) \in \mathcal{B}^{(1)}} \Bigl( \sum_{q=1}^{\omega[(n,t), \boldsymbol{r}, \boldsymbol{\theta}]} \frac{\mu[n]}{q} - \sum_{q=1}^{\omega[(n,t), \boldsymbol{r}', \boldsymbol{\theta}]} \frac{\mu[n]}{q} \Bigr) \nonumber \\ 
	& + \sum_{(n,t) \in \mathcal{B}^{(2)}} \Bigl( \sum_{q=1}^{\omega[(n,t), \boldsymbol{r}, \boldsymbol{\theta}]} \frac{\mu[n]}{q} - \sum_{q=1}^{\omega[(n,t), \boldsymbol{r}', \boldsymbol{\theta}]} \frac{\mu[n]}{q} \Bigr) \nonumber \\ 
=  & A + \sum_{(n,t) \in \mathcal{B}^{(1)}} \frac{\mu[n]}{\omega[(n,t), \boldsymbol{r}, \boldsymbol{\theta}]} - \sum_{(n,t) \in \mathcal{B}^{(2)}} \frac{\mu[n]}{\omega[(n,t), \boldsymbol{r}', \boldsymbol{\theta}]}  \nonumber \\ 
=  & A + \hspace{-0.4cm} \sum_{(n,t) \in \mathcal{B}^{(1)} \cup \mathcal{B}^{(3)}} \frac{\mu[n]}{\omega[(n,t), \boldsymbol{r}, \boldsymbol{\theta}]} - \hspace{-0.3cm} \sum_{(n,t) \in \mathcal{B}^{(2)} \cup \mathcal{B}^{(3)}} \frac{\mu[n]}{\omega[(n,t), \boldsymbol{r}', \boldsymbol{\theta}]}  \nonumber \\
=  & A + \sum_{(n,t) \in \mc{V}(\bs{r}_i)} \frac{\mu[n]}{\omega[(n,t), \boldsymbol{r}, \boldsymbol{\theta}]} - \sum_{(n,t) \in \mc{V}(\bs{r}_i')} \frac{\mu[n]}{\omega[(n,t), \boldsymbol{r}', \boldsymbol{\theta}]} \nonumber \\ 
=  & U_i(\boldsymbol{r}_i, \boldsymbol{r}_{-i}, \boldsymbol{\theta}) - U_i(\boldsymbol{r}_i', \boldsymbol{r}_{-i}, \boldsymbol{\theta}). 
\end{align}
Here, the first equality is due to the definition in \eqref{equ:potential_deterministic}.
  The second equality is due to $\bs{r}_j' = \bs{r}_j$ for $j \neq i$ and $\mathcal{B}^{(1)} \cup \mathcal{B}^{(2)} \cup \mathcal{B}^{(3)} \cup \mathcal{B}^{(4)} = \mc{N} \times \mc{T}$.
  The third equality is due to the fact that
\begin{equation}
	 \sum_{q=1}^{\omega[(n,t), \boldsymbol{r}, \boldsymbol{\theta}]} \frac{\mu[n]}{q} - \sum_{q=1}^{\omega[(n,t), \boldsymbol{r}', \boldsymbol{\theta}]} \frac{\mu[n]}{q} = 0, \text{ for } (n,t) \in \mathcal{B}^{(3)} \cup \mathcal{B}^{(4)}.
\end{equation}
  The fourth equality is due to the algebraic manipulation based on \eqref{equ:change_congestion}.
	The fifth equality is due to $\omega[(n,t), \boldsymbol{r}, \boldsymbol{\theta}] = \omega[(n,t), \boldsymbol{r}', \boldsymbol{\theta}]$ if $(n,t) \in \mc{B}^{(3)}$ from \eqref{equ:change_congestion}.
	The sixth equality is due to $\mc{V}(\bs{r}_i) = \mathcal{B}^{(1)} \cup \mathcal{B}^{(3)}$ and $\mc{V}(\bs{r}_i') = \mathcal{B}^{(2)} \cup \mathcal{B}^{(3)}$.
	The last equality is due to the definition in \eqref{equ:utility_route}.  \hfill \IEEEQED 

\subsection{Proof of Theorem \ref{thm:fip}} \label{app:fip}

  First, we define a function 
\begin{equation} \label{equ:potential_random}
	\Psi(\bs{s}) \triangleq \sum_{\boldsymbol{\theta} \in \Theta} \Phi \bigl( \boldsymbol{s}(\boldsymbol{\theta}), \boldsymbol{\theta} \bigr) p(\boldsymbol{\theta}).
\end{equation}
  We can show that
\begin{equation} 
\begin{split} 
EU_i(\bs{s}) - EU_i(\bs{s}') = \sum_{\boldsymbol{\theta} \in \Theta} \Bigl( U_i \bigl( \boldsymbol{s}(\boldsymbol{\theta}), \boldsymbol{\theta} \bigr) - U_i \bigl( \boldsymbol{s}'(\boldsymbol{\theta}), \boldsymbol{\theta} \bigr) \Bigr) p(\boldsymbol{\theta}) \\ 
= \sum_{\boldsymbol{\theta} \in \Theta} \Bigl( \Phi \bigl( \boldsymbol{s}(\boldsymbol{\theta}), \boldsymbol{\theta} \bigr) - \Phi \bigl( \boldsymbol{s}'(\boldsymbol{\theta}), \boldsymbol{\theta} \bigr) \Bigr) p(\boldsymbol{\theta}) 
= \Psi(\bs{s}) - \Psi(\bs{s}'),
\end{split} 
\end{equation}
%
%
where the first equality is due to the definition in \eqref{equ:expectedutility}.
  The second equality is due to \eqref{equ:potential_deterministic_def} and Theorem \ref{thm:bpg}.
  Thus, $\Psi(\bs{s})$ is the \emph{potential function} of game $\Omega$.
  From \cite{monderer_pg96}, every finite game with a potential function has the FIP. 	\hfill \IEEEQED 
	
\subsection{Proof of Theorem \ref{thm:complexity}} \label{app:complexity}

  As illustrated in Fig. \ref{fig:route}, in a network-time graph, the total number of nodes $V = NT$. In the extreme case that every pair of nodes is connected by an edge, the total number of edges $E \approx V^2 = N^2 T^2$.
	In computing the best response update for each type $\bs{\theta}_i$ of user $i$, due to the throughput and switching cost terms with positive and negative impacts in the utility function in \eqref{equ:utility_route}, respectively, we need to apply a shortest path algorithm that can handle both the positive and negative edge costs \cite{weiss_ds97}. It includes the Bellman-Ford algorithm, which has a computational complexity of $\mc{O}(VE) = \mc{O}(N^3 T^3)$. 
	Overall, since user $i$ has $|\Theta_i|$ possible types for his strategy, each best response update requires $\mc{O}(|\Theta_i| N^3 T^3)$ time. \hfill \IEEEQED 

%

\bibliographystyle{IEEEtran}
\bibliography{IEEEabrv,mybibfile}

\begin{IEEEbiography}
[{\includegraphics[width=1in,height=1.25in,clip,keepaspectratio]{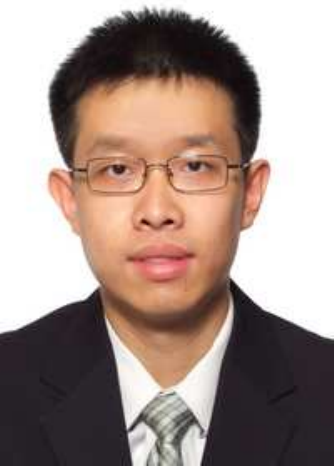}}]
{Man Hon Cheung} received the B.Eng. and M.Phil. degrees in Information Engineering from the Chinese University of Hong Kong (CUHK) in 2005 and 2007, respectively, and the Ph.D. degree in Electrical and Computer Engineering from the University of British Columbia (UBC) in 2012. Currently, he is a postdoctoral fellow in the Department of Electrical and Computer Engineering at the University of Macau. He worked as a postdoctoral fellow in the Department of Information Engineering in CUHK. He received the IEEE Student Travel Grant for attending {\it IEEE ICC 2009}. He was awarded the Graduate Student International Research Mobility Award by UBC, and the Global Scholarship Programme for Research Excellence by CUHK. He serves as a Technical Program Committee member in {\it IEEE ICC}, {\it Globecom}, {\it WCNC}, and {\it WiOpt}. His research interests include the design and analysis of wireless network protocols using optimization theory, game theory, and dynamic programming, with current focus on mobile data offloading, mobile crowdsensing, and network economics.
\end{IEEEbiography}

\begin{IEEEbiography}[{\includegraphics[width=1in,height=1.25in,clip,keepaspectratio]{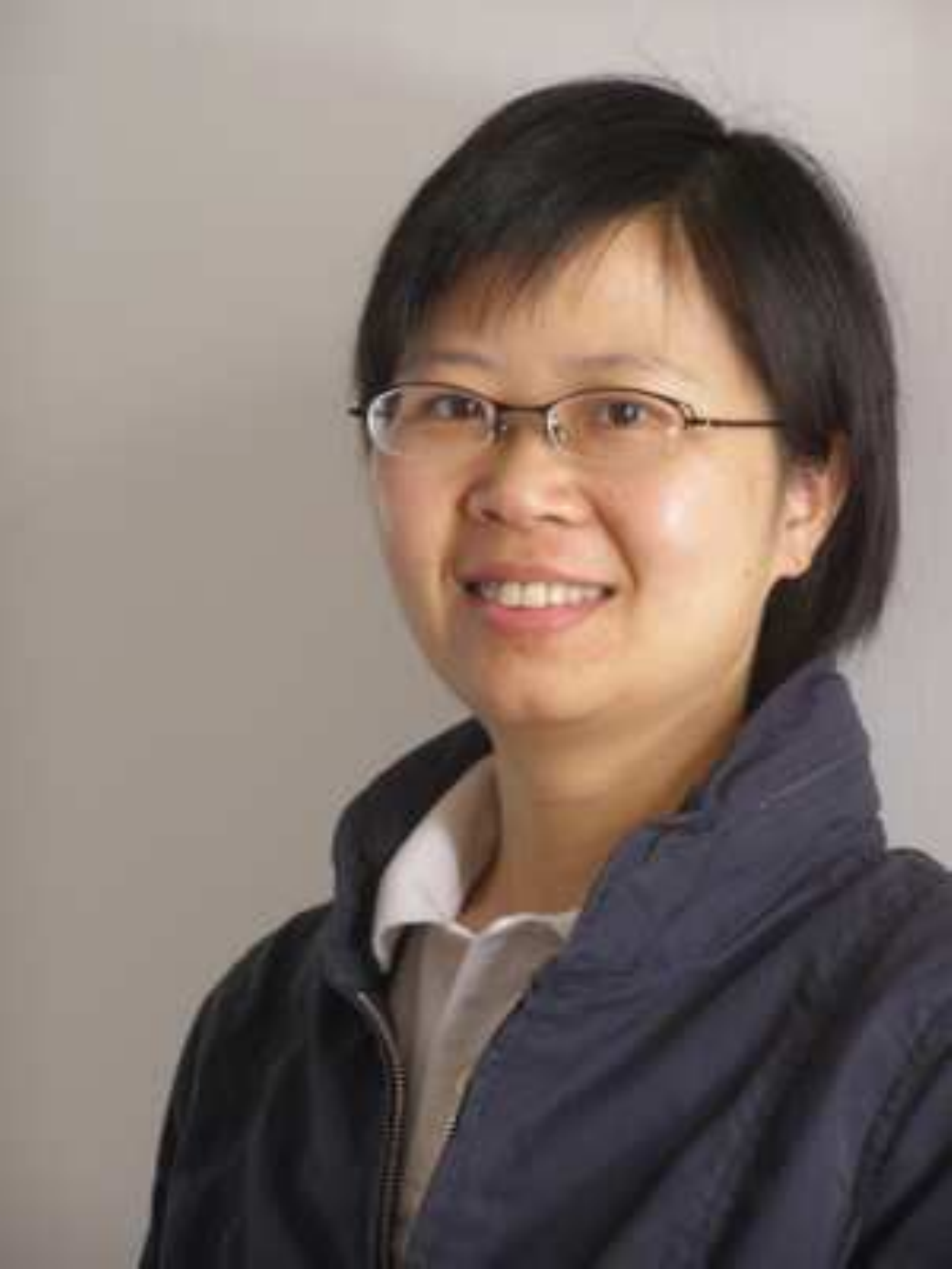}}]
{Fen Hou}(M'10) is an Assistant Professor in the Department of Electrical and Computer Engineering at the University of Macau. She received the Ph.D. degree in electrical and computer engineering from the University of Waterloo, Waterloo, Canada, in 2008. She worked as a postdoctoral fellow in the Electrical and Computer Engineering at the University of Waterloo and in the Department of Information Engineering at the Chinese University of Hong Kong from 2008 to 2009 and from 2009 to 2011, respectively. Her research interests include resource allocation and scheduling in broadband wireless networks, protocol design and QoS provisioning for multimedia communications in broadband wireless networks, Mechanism design and optimal user behavior in mobile crowd sensing networks and mobile data offloading. She is the recipient of IEEE GLOBECOM Best Paper Award in 2010 and the Distinguished Service Award in IEEE MMTC in 2011. Dr. Fen Hou served as the co-chair in ICCS 2014 Special Session on Economic Theory and Communication Networks, INFOCOM 2014 Workshop on Green Cognitive Communications and Computing Networks (GCCCN), IEEE Globecom Workshop on Cloud Computing System, Networks, and Application (CCSNA) 2013 and 2014, ICCC 2015 Selected Topics in Communications Symposium, and ICC 2016 Communication Software Services and Multimedia Application Symposium, respectively. She currently serves as the vice-chair (Asia) in IEEE ComSoc Multimedia Communications Technical Committee (MMTC) and an associate editor for IET Communications as well.
\end{IEEEbiography}

\begin{IEEEbiography}[{\includegraphics[width=1in,height=1.25in,clip,keepaspectratio]{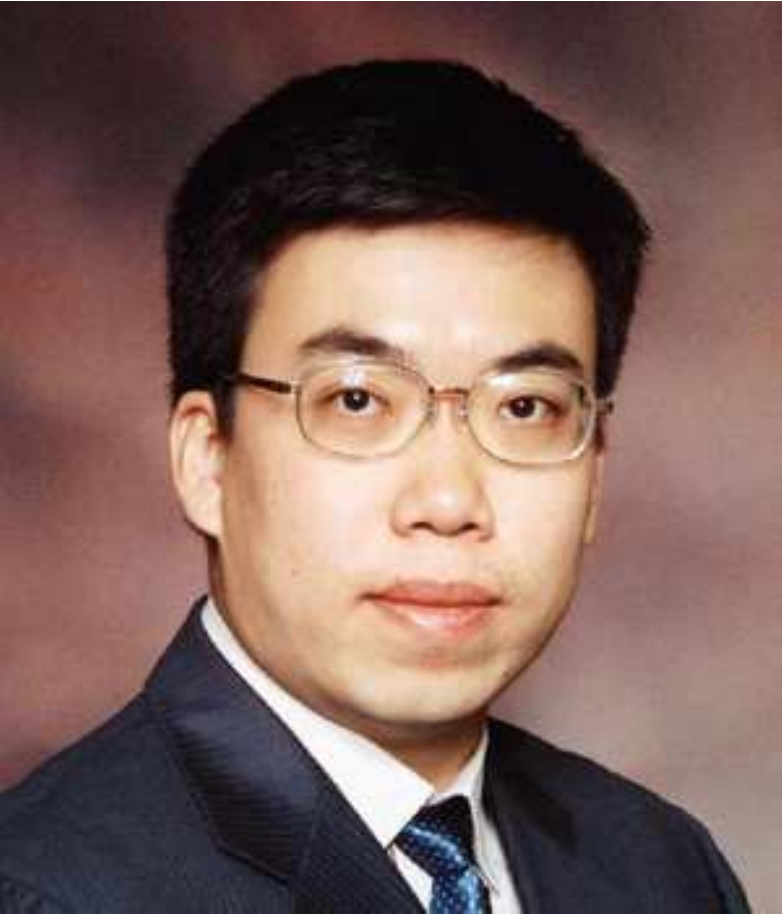}}]
{Jianwei Huang} (S'01-M'06-SM'11-F'16) is an Associate Professor and Director of the Network Communications and Economics Lab (ncel.ie.cuhk.edu.hk), in the Department of Information Engineering at the Chinese University of Hong Kong. He received the Ph.D. degree from Northwestern University in 2005, and worked as a Postdoc Research Associate at Princeton University during 2005-2007. Dr. Huang is the co-recipient of 8 Best Paper Awards, including IEEE Marconi Prize Paper Award in Wireless Communications in 2011. He has co-authored six books, including the textbook on ``Wireless Network Pricing.'' He received the CUHK Young Researcher Award in 2014 and IEEE ComSoc Asia-Pacific Outstanding Young Researcher Award in 2009. Dr. Huang has served as an Associate Editor of IEEE/ACM Transactions on Networking, IEEE Transactions on Cognitive Communications and Networking, IEEE Transactions on Wireless Communications, and IEEE Journal on Selected Areas in Communications - Cognitive Radio Series. He has served as the Chair of IEEE ComSoc Cognitive Network Technical Committee and Multimedia Communications Technical Committee. He is an IEEE Fellow, a Distinguished Lecturer of IEEE Communications Society, and a Thomson Reuters Highly Cited Researcher in Computer Science.
\end{IEEEbiography}

\begin{IEEEbiography}[{\includegraphics[width=1in,clip,keepaspectratio]{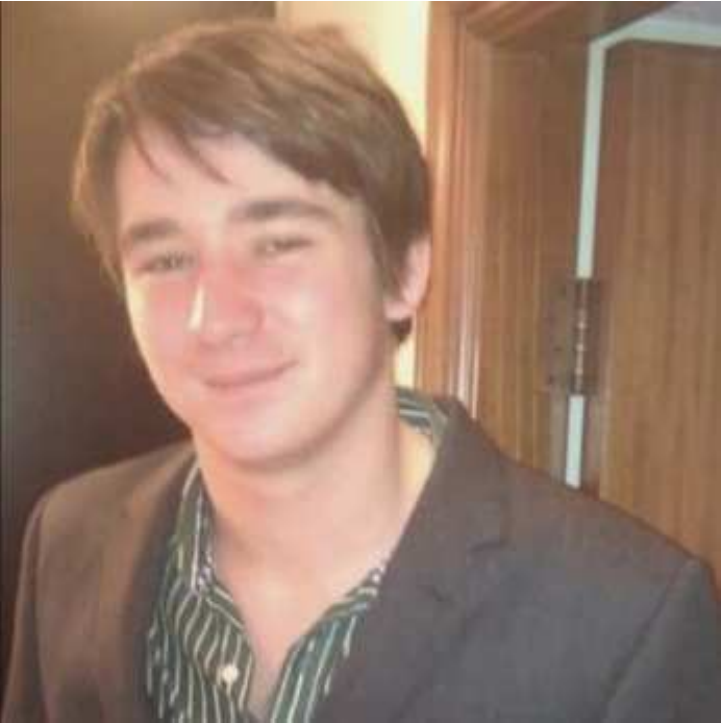}}]
{Richard Southwell} did his BSc in Theoretical Physics at the University of York, MSc in Mathematics at the University of York, and Ph.D. in Mathematics at the University of Sheffield. After working as a research associate in the amorphous computing project, he moved to Hong Kong to work as a researcher at the Network Communications and Economics Lab (NCEL) in the Information Engineering Department at the Chinese University of Hong Kong. Later he became an Assistant Professor at Institute for Interdisciplinary Information Sciences (IIIS) in Tsinghua University, Beijing. Then he moved back to Hong Kong and again worked as a researcher in NCEL, as well as at the Department of Management Science in the City University of Hong Kong. He is currently working at the York Centre for Complex Systems Analysis, in connection with the Department of Mathematics. His research interests include graph theory, game theory, complex systems, projective geometry, topology, and dynamics. Currently, he is using partial differential based equations to model marine ecology.
\end{IEEEbiography}

\end{document}